# Validation of HRV Studio: A Transparent and Quality-Control-Aware Platform for Heart Rate Variability Analysis

**Author list：**

Cyrus Mexon Evrard Djindot, Faliang Liu, Sylvain Laborde, Yinjia Zhang, Jessie Chen, Ming Li, Congrong Wang*, Weixiong Rao*, Qinpei Zhao*

# Abstract

**Background and Objective:** Reproducibility of heart rate variability (HRV) analysis is heavily hindered by "black-box" commercial platforms and complex code-based libraries, coupled with extreme sensitivity to pre-processing conventions. To bridge this gap, we developed and validated HRV Studio—an open-source PyQt6-based desktop application that integrates an interactive, audited graphical user interface (GUI) with a transparent, automated quality-control (QC) layer to support reproducible physiological research.

**Methods:** We validated the platform using a rigorous, six-stage framework: (1) large-scale numerical agreement against NeuroKit2 using normal-sinus-rhythm recordings; (2) targeted benchmarking against Kubios on a curated 44-recording subset; (3) alternative spectral-method comparison (Welch, FFT, autoregressive); (4) synthetic robustness testing under controlled noise, missed beats, and ectopy; (5) recording-duration sensitivity analysis (30 s to 10 min); and (6) software stress-testing using the MIT-BIH Arrhythmia Database to evaluate automated QC warning performance.

**Results:** Under harmonized short-term segment-linear settings, HRV Studio demonstrated strong agreement with NeuroKit2 across complementary HRV domains. In the primary five-minute comparison, median relative errors were 1.35% for $LF$, 0.18% for $HF$, 1.41% for $LF/HF$, 0.42% for $LF_{nu}$, and 0.81% for $HF_{nu}$, while total power (13.57%) and $VLF$ (37.79%) remained more convention-sensitive. Time-domain indices showed excellent agreement for RMSSD and SDNN (median relative errors <0.01%), while nonlinear Poincaré descriptors SD1 and SD2 also demonstrated high consistency under matched conditions. Extended ten-minute analyses used for sensitivity assessment showed the same overall metric-dependent pattern, with generally lower disagreement for convention-sensitive spectral outputs. Targeted Kubios benchmarking after NN-sequence harmonization demonstrated near-identical agreement for time-domain and nonlinear indices and strong agreement for most frequency-domain measures, with median relative errors below 6% for $LF$, $HF$, total power, $LF/HF$, $LF_{nu}$, and $HF_{nu}$; $VLF$ retained the largest residual error (14.01%). An additional matched Smoothness Priors sensitivity analysis demonstrated substantially closer agreement for autoregressive spectral estimates than for FFT-derived estimates. Synthetic perturbations and arrhythmia stress-testing yielded 100% numerical stability and consistently surfaced automated QC warnings under degraded or non-normal rhythm conditions. Welch PSD estimation demonstrated greater methodological stability than FFT and autoregressive alternatives, while duration analysis showed substantially greater instability in ultra-short recordings and persistent, non-monotonic sensitivity across short-term durations.

**Conclusions:** HRV Studio provides a benchmarked and validated, highly transparent alternative for physiological research. Its primary strengths lie in its reproducible preprocessing framework, quality-control-aware workflow, and validated computation across complementary HRV domains, including time-domain, frequency-domain, and nonlinear indices. The validation demonstrated high numerical consistency for commonly used time-domain measures such as RMSSD and SDNN, nonlinear Poincaré descriptors, and most frequency-domain measures when effective NN sequences, preprocessing settings, and spectral conventions were explicitly harmonized, while identifying VLF as particularly sensitive to methodological choices. Crucially, while the software shows high computational robustness under degraded and arrhythmic signals, these evaluations serve strictly as software stress tests rather than clinical or diagnostic validation.

**Index Terms—** heart rate variability (HRV), signal processing, physiological computing, spectral analysis, software validation, quality control, reproducibility.

# 1. Introduction

Heart rate variability (HRV), defined as the variation in timing between successive cardiac intervals, is a widely used framework for quantifying autonomic and cardiorespiratory dynamics in biomedical engineering, physiology, and clinical research [1], [2], [3]. HRV analysis has been applied across domains including cardiovascular physiology, stress and recovery assessment, sleep research, exercise science, neurophysiology, and disease monitoring because interval dynamics may reflect interactions among autonomic, respiratory, and regulatory processes [1], [2], [4], [5], [6]. Its breadth of use has been theorized to stem from its ability to index overall self-regulation mechanisms related to cardiac vagal activity, as outlined in the neurovisceral integration model [7] and the vagal tank theory [8]. Standard HRV analysis typically includes time-domain, frequency-domain, and nonlinear measures derived from normal-to-normal (NN) interval sequences, with metrics such as the standard deviation of normal-to-normal intervals (SDNN), root mean square of successive differences (RMSSD), low-frequency power ($LF$), high-frequency power ($HF$), and the low-frequency to high-frequency power ratio ($LF/HF$) frequently reported across experimental settings [1], [2].

Despite widespread use, HRV estimation is highly sensitive to methodological conventions. Reported values can be materially affected by preprocessing decisions, artifact correction, interpolation rate, detrending strategy, spectral estimation method, frequency-band implementation, and recording duration [1], [3], [9], [10], [11], [12], [13], [14]. In particular, frequency-domain estimates may vary substantially across software frameworks because differences in power spectral density (PSD) computation, direct-current (DC) handling, frequency-bin integration, and segment processing can influence low-frequency power estimates even when identical physiological recordings are analyzed [9], [15]. Such variability complicates cross-platform reproducibility and can obscure whether disagreement arises from methodological assumptions or underlying physiological signal content [16]. These challenges motivate the need for transparent and reproducible HRV workflows in which preprocessing assumptions, spectral conventions, and quality-control (QC) indicators are explicitly surfaced rather than hidden from the user. Recent methodological literature has emphasized reproducibility, preprocessing transparency, and standardized HRV reporting to improve interpretability and comparability across studies [2], [5], [17], [18]. In this context, software validation should extend beyond numerical output alone and include explicit assessment of preprocessing behavior, robustness under degraded signal conditions, and visibility of conditions that may reduce interpretability. Contemporary HRV analysis frameworks generally compute metrics spanning time-domain, frequency-domain, and nonlinear representations of interval variability [2]. Time-domain metrics, such as SDNN and RMSSD, summarize overall and short-term variability, whereas frequency-domain measures estimate power within physiologically defined spectral bands including very-low-frequency ($VLF$), $LF$, and $HF$ components [1], [2]. Nonlinear methods, including Poincaré analysis, entropy measures, and detrended fluctuation analysis (DFA), further characterize signal complexity and temporal organization [2], [19].

Frequency-domain estimation remains particularly sensitive to methodological implementation. Common PSD approaches include Welch spectral estimation, fast Fourier transform (FFT)- based periodograms, and autoregressive (AR) modeling, each involving distinct assumptions regarding segmentation, smoothing, spectral resolution, and noise behavior [9], [20], [21]. Welch methods are widely used because they provide stable PSD estimates through segmented averaging, whereas FFT and AR approaches may exhibit greater sensitivity to detrending conventions, model order, or low-frequency spectral behavior [9], [15], [22]. Consequently, seemingly small implementation differences can produce materially different frequency-domain outputs despite identical RR interval data. Several open-source and commercial frameworks are commonly used for HRV analysis. NeuroKit2, a widely adopted Python toolbox for neurophysiological signal processing, provides integrated time-domain, frequency-domain, and nonlinear HRV estimation together with transparent parameterization and reproducible workflows [4]. Kubios HRV, based on a graphical user interface, is another extensively used platform and has become a common benchmark in methodological HRV studies because of its mature preprocessing and spectral-analysis implementation [23]. However, cross-platform comparison remains challenging because software packages frequently differ in detrending defaults, interpolation behavior, PSD integration conventions, and artifact correction strategies [24], [25]. Methodological recommendations also emphasize the importance of recording duration. The 1996 Task Force standards recommended approximately 5 min recordings for short-term frequency-domain HRV analysis, while cautioning against interpretation of VLF from short recordings because of limited low-frequency resolution [1]. Regarding RMSSD, the most commonly used time-domain index, a minimum recording duration of one minute is recommended to reliably index cardiac vagal activity [3]. More recent studies similarly report reduced reliability and increased instability in ultra-short recordings, particularly for frequency-domain measures and ratio-based outputs [10], [26], [27]. Consequently, recording duration remains an important determinant of interpretability in HRV validation studies. Finally, recordings containing substantial arrhythmia or ectopic activity are generally unsuitable as primary material for standard HRV agreement validation because rhythm irregularity can fundamentally alter interval dynamics and complicate interpretation of conventional HRV metrics [1], [28]. Such recordings are therefore more appropriately treated as robustness and preprocessing stress-test conditions rather than evidence of physiological or methodological equivalence.

HRV Studio was developed as a transparent, reproducible, and quality-control-aware HRV analysis platform intended to support research workflows rather than function as a diagnostic system. The software integrates interval preprocessing, artifact correction, spectral analysis, nonlinear HRV estimation, visualization, and export within a unified Python/PyQt6 desktop environment while exposing preprocessing assumptions and QC diagnostics to the user. As an open-source, GUI-based platform designed for modular integration into physiological monitoring pipelines—including HRV-based glucose prediction and stress-monitoring applications—HRV Studio addresses a gap not covered by existing code-based libraries or closed-source commercial tools.

The objective of the present study was to evaluate HRV Studio through a staged software-validation framework emphasizing agreement under matched preprocessing and spectral settings, robustness to controlled signal perturbations, recording-duration sensitivity, quality-control transparency and warning behavior, and targeted benchmark comparison under constrained matched conditions. Validation included large-scale comparison against NeuroKit2 using normal-sinus-rhythm recordings, a manually reviewed Kubios benchmark subset, methodological evaluation of FFT and AR spectral approaches, synthetic perturbation testing, and arrhythmia-focused robustness assessment. Recordings from the MIT-BIH Arrhythmia Database (MITDB) were used exclusively as engineering robustness and quality-control stress-test data to evaluate preprocessing behavior, numerical stability, and warning performance under heterogeneous rhythm conditions.

# 2. Methods

## 2.1 Software Architecture and Implementation

### System Overview

HRV Studio was implemented as a Python-based desktop application and modular analysis library for transparent heart rate variability (HRV) analysis. The graphical user interface (GUI) was developed using PyQt6 and integrates RR interval preprocessing, HRV metric computation, visualization, quality-control diagnostics, and export functionality within a unified desktop workflow (Figure 1). The source code and software implementation of HRV Studio are publicly available at https://github.com/CyrusMexon/HRV-Analysis_TJU. The software is intended to support reproducible and quality-controlled HRV research by exposing preprocessing settings, spectral conventions, and diagnostic outputs rather than abstracting them from the user. The software architecture follows a layered modular design consisting of (i) data ingestion, (ii) preprocessing and quality control, (iii) HRV computation, (iv) visualization, (v) export and audit logging, and (vi) validation support modules [29]. This organization separates signal preparation from metric estimation and reporting while enabling reproducible analysis workflows and method-specific diagnostics (Figure 2). The central analysis pathway proceeds from file import and interval loading through preprocessing and optional correction, followed by unified computation of time-domain, frequency-domain, nonlinear, and respiratory-derived metrics. Computed outputs are subsequently integrated with quality indicators, visualization tools, and export mechanisms to support transparent downstream interpretation.

### Input Data Handling

HRV Studio supports multiple interval- and waveform-based input formats commonly used in biomedical and physiological research. Implemented readers include comma-separated value (CSV) and plain text (TXT) interval files, European Data Format (EDF) waveforms, Polar HRM files, Garmin FIT files, Suunto SML files, Movesense-style JSON data, and Biopac ACQ recordings when compatible dependencies are available. Input files may contain RR interval sequences, pulse-to-pulse intervals, electrocardiography (ECG), photoplethysmography (PPG), and respiratory signals.

Internally, imported data are represented using structured containers that preserve interval data, waveform signals, source metadata, and preprocessing information. When explicit interval sequences are unavailable, HRV Studio can derive interval series from ECG or PPG waveforms using NeuroKit2-based signal processing methods. However, the present validation package primarily evaluated downstream NN interval processing and HRV metric agreement rather than independent validation of peak-detection accuracy from raw physiological waveforms. Accordingly, waveform-derived interval extraction is described as an implemented capability but was not treated as a primary validation endpoint in the present study.

### Visualization, Editing, and Reproducibility Features

The GUI provides an integrated workflow for data loading, parameter configuration, threaded analysis execution, signal visualization, manual editing, and export (Figure 1) [30], [31]. Threaded processing was implemented to prevent blocking of the user interface during computationally intensive analyses. Users can configure preprocessing and spectral-analysis settings, review warnings and quality indicators, and inspect computed HRV outputs within a single analysis environment. Visualization tools include RR tachograms, heart-rate distribution displays, power spectral density (PSD) plots, frequency-band summaries, Poincaré plots, detrended fluctuation analysis (DFA) visualizations, and quality-assessment summaries. Manual beat-editing functionality supports interval insertion, deletion, movement, interpolation, undo operations, and reanalysis, thereby allowing investigator-supervised correction of problematic recordings while preserving transparency of editing actions. To support reproducibility and downstream reporting, HRV Studio implements export functionality for PDF reports, comma-separated metric tables, SPSS-compatible CSV files, and structured JavaScript Object Notation (JSON) audit trails. Exported outputs can include preprocessing summaries, analysis parameters, quality indicators, and manual-edit histories. In addition, validation workflows generate run-level reproducibility artifacts, including structured metadata, diagnostics, comparison tables, and analysis summaries to facilitate traceable benchmarking and repeatability.

**A. Main analysis interface**

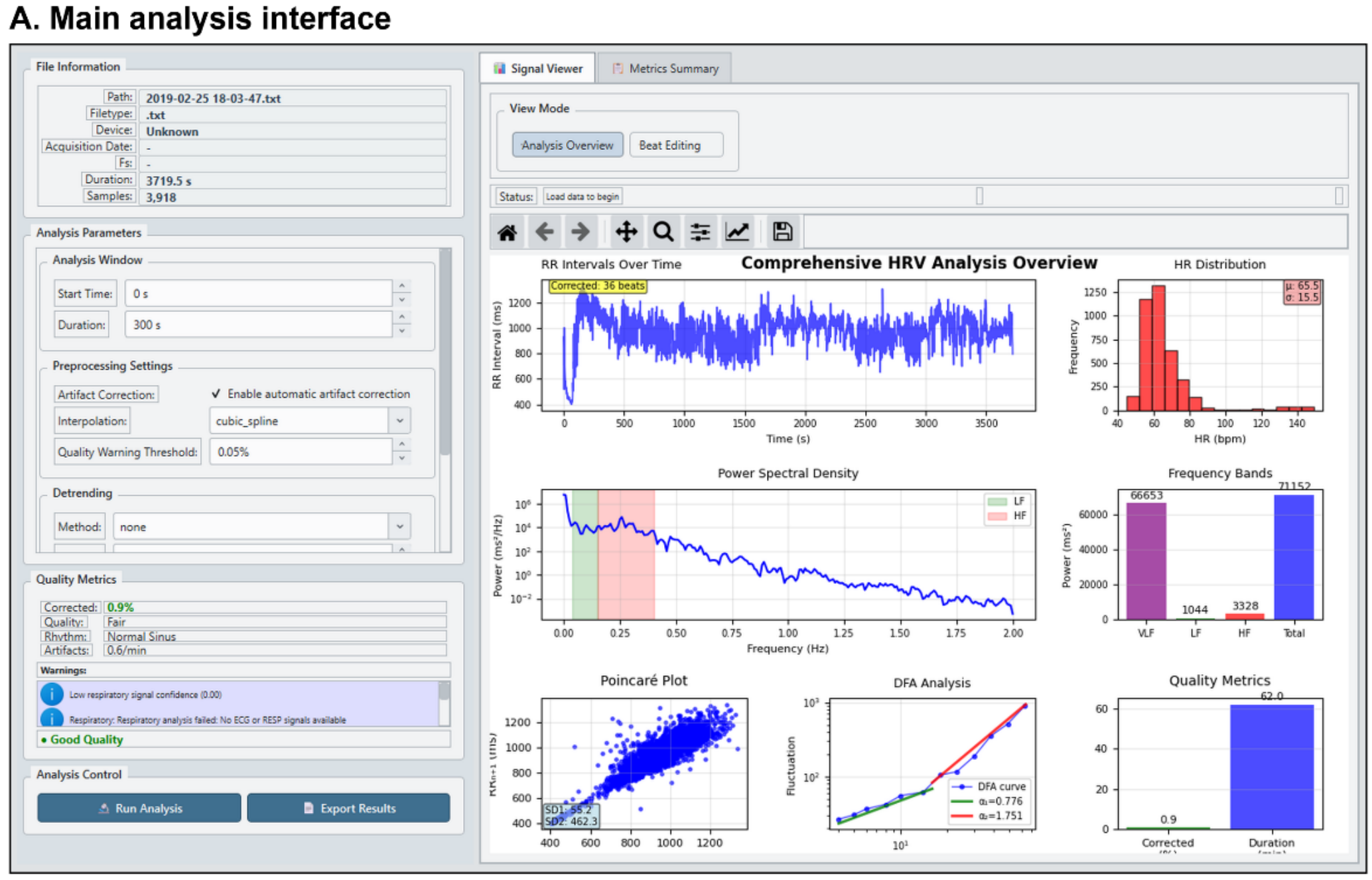

**B. Beat editing interface**

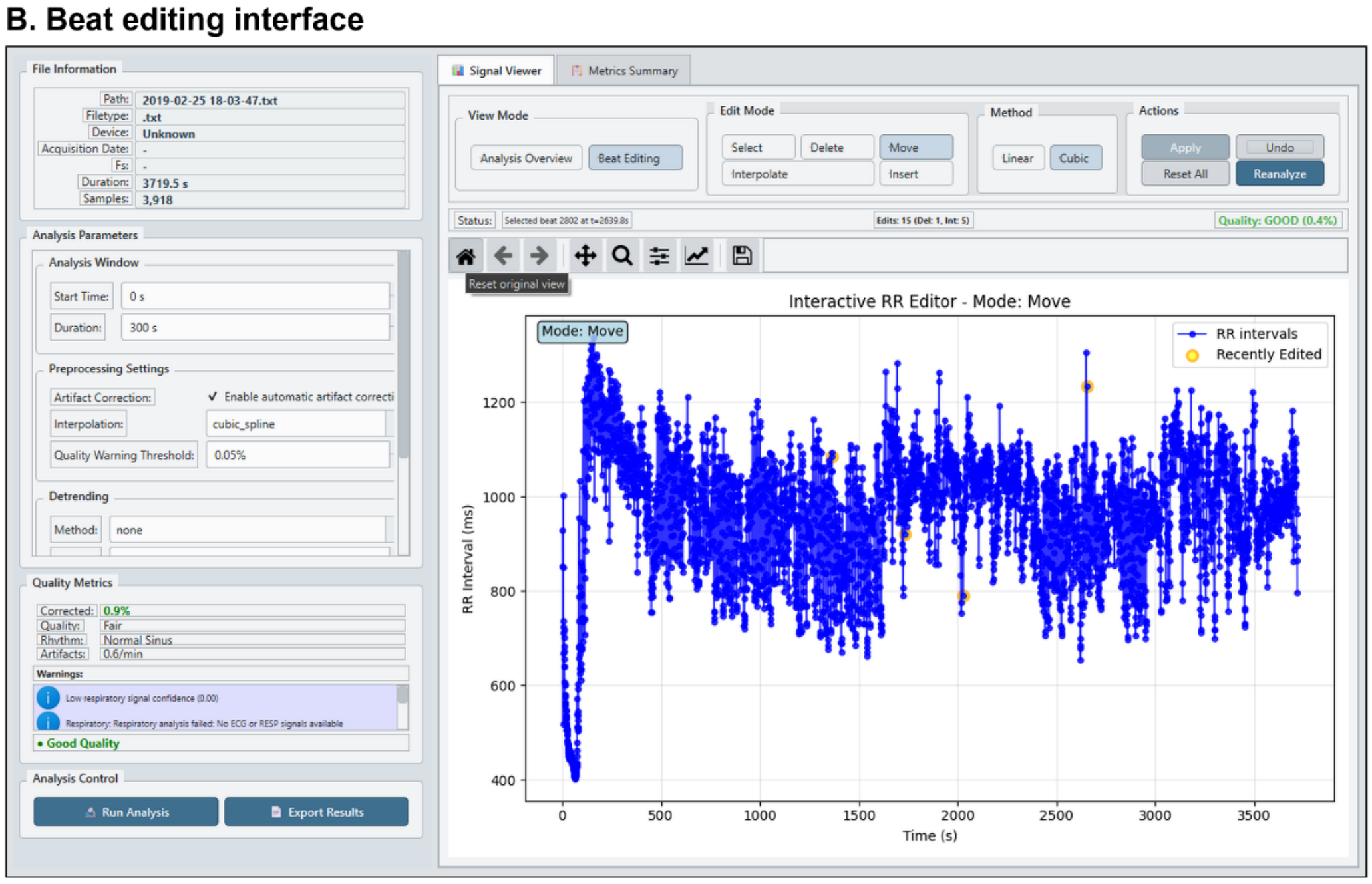

*Figure 1. HRV Studio graphical user interface. The software provides integrated RR interval preprocessing, time-domain, frequency-domain, and nonlinear HRV analysis, together with visualization, quality-control diagnostics, and export functionality.*

## 2.2 RR-to-NN Interval Preprocessing and Quality Control

The preprocessing pipeline was designed to support transparent, reproducible, and quality-controlled HRV analysis while minimizing nonphysiological signal distortions. Rather than treating metric generation as independent of signal preparation, HRV Studio integrates preprocessing diagnostics and warning mechanisms into the analysis workflow to facilitate interpretation of potentially unstable or convention-sensitive outputs (Figure 2).

### RR Interval Validation and Artifact Handling

Throughout this manuscript, RR intervals refer to the raw beat-to-beat interval sequences imported or extracted from physiological recordings. Following artifact detection, correction, interpolation where applicable, and exclusion of non-sinus beats, the resulting cleaned sequences are referred to as normal-to-normal (NN) intervals in accordance with the Task Force recommendations.

Input RR interval sequences underwent validation to ensure physiologically plausible analysis conditions prior to preprocessing and HRV analysis. Unit normalization was applied when necessary to harmonize interval representations, and invalid observations, including non-finite or nonpositive intervals, were excluded before downstream processing.

Artifact detection was performed to identify interval patterns consistent with missed beats, ectopic intervals, extra beats, and abrupt timing discontinuities [32]. Detection procedures combined interval-threshold screening with local variability criteria to identify potentially nonphysiological patterns requiring review or correction. When correction was enabled, HRV Studio applied interpolation-based reconstruction strategies, including cubic spline interpolation where appropriate, while preserving transparency regarding preprocessing actions and modified intervals [33]. To support auditability and downstream quality assessment, preprocessing retained both the original RR interval sequence and the resulting NN interval sequence. Artifact counts, correction summaries, and preprocessing metadata were propagated to subsequent analysis and reporting stages to enable explicit interpretation of corrected recordings.

### Quality-Control Diagnostics and Warning System

HRV Studio incorporates a quality-control layer intended to support transparent interpretation rather than guarantee physiological validity. Diagnostic outputs include signal-quality indicators, duration-related warnings, frequency-band sufficiency checks, rhythm-irregularity indicators, and method-specific spectral diagnostics (Figure 2). Frequency-domain analyses additionally report characteristics relevant to interpretation, including effective Welch segmentation behavior, spectral-band coverage, and selected spectral-consistency diagnostics. Warnings are surfaced to identify potentially unreliable analysis conditions, including insufficient frequency-band resolution, unstable very-low-frequency ( $VLF$ ) estimation, excessive artifact burden or correction dependence, and irregular rhythm structure. These outputs are intended to function as screening and interpretability aids rather than definitive indicators of physiological validity. Consequently, finite numerical output was not assumed to imply reliable physiological interpretation in the absence of supporting signal quality or methodological sufficiency.

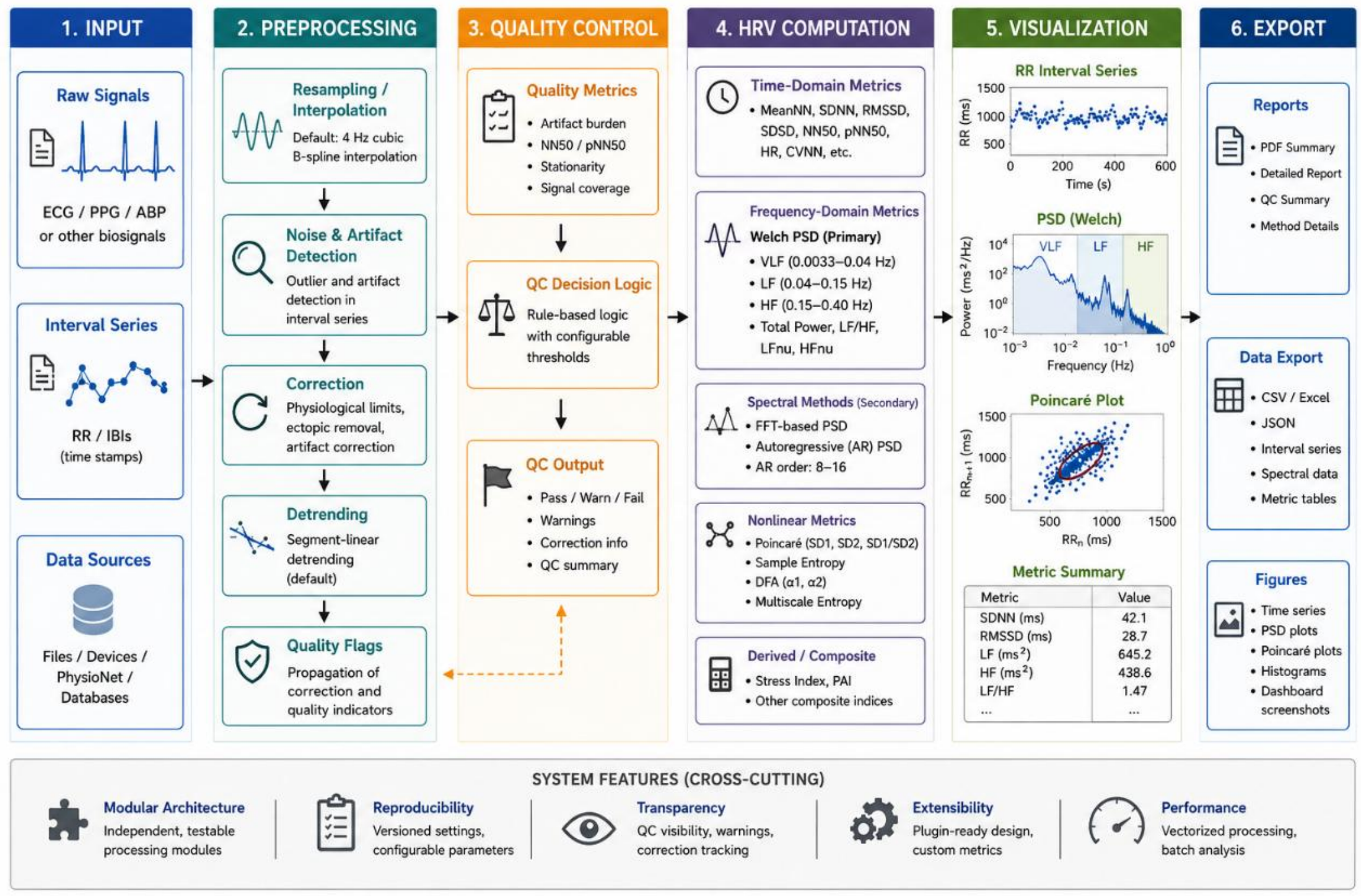

***Figure 2.*** HRV Studio processing architecture. Physiological signals and interval series undergo preprocessing, quality-control assessment, HRV computation, visualization, and export through a modular analysis pipeline. Abbreviations: ECG, electrocardiography; PPG, photoplethysmography; ABP, arterial blood pressure; RR, R–R interval; IBI, inter-beat interval; HRV, heart rate variability; QC, quality control; NN, normal-to-normal interval; RMSSD, root mean square of successive differences; SDNN, standard deviation of normal-to-normal intervals; pNN50, percentage of successive normal-to-normal intervals differing by more than 50 ms; PSD, power spectral density; VLF, very-low-frequency; LF, low-frequency; HF, high-frequency; LF/HF, low-frequency to high-frequency power ratio; LFnu/HFnu, normalized low-frequency/high-frequency power; FFT, fast Fourier transform; AR, autoregressive; SD1/SD2, Poincaré plot standard deviations; DFA, detrended fluctuation analysis; CSV, comma-separated values; JSON, JavaScript Object Notation; PDF, portable document format.

## 2.3 HRV Metric Computation

HRV Studio computes time-domain, frequency-domain, nonlinear, and respiratory-derived metrics from validated NN interval sequences produced by the preprocessing pipeline [34]. The analysis framework was designed to provide transparent computation together with method-specific diagnostics and explicit preprocessing traceability. Frequency-domain analyses were implemented using multiple spectral estimation approaches, with Welch power spectral density (PSD) estimation treated as the primary reporting method in the present validation package.

### *Time-Domain Analysis*

Time-domain HRV measures were computed from preprocessed normal-to-normal (NN) interval sequences using standard variability statistics commonly reported in physiological and biomedical HRV research. Implemented metrics included the standard deviation of NN intervals (SDNN), root mean square of successive differences (RMSSD), percentage of adjacent intervals differing by more than 50 ms (pNN50), percentage differing by more than 20 ms (pNN20), triangular interpolation of the NN interval histogram (TINN), and the HRV triangular index, together with additional descriptive variability measures including mean RR interval, heart rate, coefficient-of-variation metrics, and interval distribution statistics.

SDNN was computed as the standard deviation of NN intervals:

$$SDNN = \sqrt{\frac{1}{N-1}\sum_{i=1}^{N}\left(RR_i - \overline{RR}\right)^2} \quad (1)$$

where $RR_i$ represents the $i$-th interval and $\overline{RR}$ denotes the mean interval duration.

RMSSD was calculated to quantify short-term beat-to-beat variability:

$$RMSSD = \sqrt{\frac{1}{N-1}\sum_{i=1}^{N-1}(RR_{i+1} - RR_i)^2} \quad (2)$$

The pNN50 and pNN20 metrics were computed as:

$$pNN_x = \frac{NN_x}{N-1} \times 100 \quad (3)$$

where $NN_x$ represents the number of adjacent interval pairs differing by more than $x$ milliseconds (20 or 50 ms). These measures primarily characterize short-term variability and parasympathetic-related fluctuations, whereas SDNN reflects broader overall variability across the analyzed recording duration.

***Frequency-Domain Analysis***

Frequency-domain HRV analysis was performed using Welch PSD estimation as the primary spectral method. In addition, whole-signal fast Fourier transform (FFT) periodogram estimation and adaptive autoregressive (AR) spectral estimation were implemented as secondary methods to support methodological comparison and diagnostic evaluation. Consistent with the present validation design, Welch-derived estimates were prioritized for interpretation and reporting, while FFT and AR outputs were treated as method-dependent secondary analyses. Prior to spectral estimation, NN interval sequences were interpolated to a uniformly sampled signal. Under the primary validation settings, interpolation was performed at 4 Hz using Welch PSD estimation with a 120 s segment length, 75% overlap, and a Hann window. Detrending conventions were explicitly parameterized to support matched-setting comparisons and methodological transparency.

Frequency-domain outputs were computed using mutually exclusive physiological frequency bands under the standard HRV Studio analysis convention: ultra-low frequency ($ULF$), $0 < f < 0.003\ Hz$; very-low frequency ($VLF$), $0.003 \leq f < 0.04\ Hz$; low frequency ($LF$), $0.04 \leq f < 0.15\ Hz$; and high frequency ($HF$), $0.15 \leq f \leq 0.40\ Hz$. The direct-current component (f=0) was excluded from physiological band-power calculations, and adjacent frequency-band boundaries were assigned exclusively to prevent double counting

Band-specific spectral power was estimated by integrating the PSD over the corresponding frequency interval:

$$P_{band} = \int_{f_1}^{f_2} PSD(f)\ df \quad (4)$$

where $f_1$ and $f_2$ denote the lower and upper limits of the corresponding frequency band.

Total physiological power was computed by direct integration over $0 < f \leq 0.40\ Hz$. The $LF/HF$ ratio was calculated as:

$$LF/HF = \frac{LF}{HF} \tag{5}$$

Normalized spectral units were computed to emphasize relative autonomic balance independent of absolute spectral magnitude:

$$LF_{nu} = \frac{LF}{LF + HF} \times 100 \tag{6}$$

$$HF_{nu} = \frac{HF}{LF + HF} \times 100 \tag{7}$$

Additional spectral outputs included relative band powers and peak frequencies for the $VLF$, $LF$, and $HF$ bands. Method-specific diagnostics were retained to document effective segment behavior, frequency resolution, band sufficiency, and spectral consistency indicators. Because spectral estimates are sensitive to preprocessing conventions, detrending choices, interpolation, and PSD integration behavior, these parameters were explicitly exposed and recorded to support reproducibility.

### ***Nonlinear and Respiratory Measures***

HRV Studio additionally implements nonlinear HRV analyses intended to characterize signal complexity and temporal organization beyond conventional variability statistics. Poincaré plot measures included short-axis variability (SD1), long-axis variability (SD2), their ratio, and ellipse-based descriptors. SD1 and SD2 were computed from transformed interval differences and summarize short- and longer-term variability components, respectively. Entropy-based measures included sample entropy and approximate entropy for quantifying signal irregularity, together with multiscale entropy (MSE) analysis to characterize complexity across multiple temporal scales. Detrended fluctuation analysis (DFA) was implemented to estimate short- and long-range fractal scaling exponents ($\alpha_1$ and $\alpha_2$). Additional complexity-oriented analyses and respiratory-support modules were also implemented. Respiratory-derived measures may incorporate direct respiratory channels or ECG/PPG-derived respiratory estimation when available, including respiratory sinus arrhythmia–related information. However, respiratory analyses and nonlinear complexity measures were not central endpoints of the present validation package and are therefore described only briefly in the current study.

## 2.4 Validation Design and Comparator Framework

### Validation Overview

The validation strategy was designed as a staged evaluation framework progressing from controlled agreement assessment toward robustness and stress testing. Rather than relying on a single comparator or dataset, validation was structured to examine agreement under matched analysis settings, sensitivity to methodological conventions, operational robustness under degraded signal conditions, and transparency of quality-control behavior (Figure 3). Validation datasets included publicly available RR interval and arrhythmia resources selected to support distinct methodological objectives. Primary agreement analyses were based on normal-sinus-rhythm recordings obtained from the **MIT-BIH Normal Sinus Rhythm Database** (NSRDB) and complementary PhysioNet RR interval resources, which provided relatively clean interval sequences appropriate for matched-setting HRV agreement assessment. These recordings served as the principal source for large-scale NeuroKit2 comparison, duration sensitivity analyses, and spectral-method evaluation because they minimized confounding effects from rhythm abnormalities and enabled controlled evaluation of preprocessing and power spectral density (PSD) conventions. A separate set of predefined analysis segments derived from the **MIT-BIH Arrhythmia Database** (MITDB) was used exclusively for robustness and quality-control stress testing (Table 2). These recordings contain rhythm irregularities, ectopic activity, and clinically heterogeneous patterns that can substantially alter conventional HRV interpretation. Consequently, arrhythmia data were not treated as primary validation material for standard HRV agreement analyses and were instead used to evaluate numerical stability, warning behavior, preprocessing sensitivity, and operational robustness under non-normal rhythm conditions.

Validation phases were performed sequentially:

**1. Large-scale NeuroKit2 agreement validation**, designed to assess metric agreement under matched preprocessing and spectral-analysis conditions using large-scale PhysioNet-derived RR interval recordings.

**2. Kubios benchmark subset comparison**, designed as a targeted benchmark against manually exported Kubios reports using quality-controlled and manually reviewed recordings.

**3. FFT and autoregressive (AR) spectral method evaluation**, performed to assess methodological consistency and support Welch PSD as the primary reporting approach.

**4. Synthetic robustness and signal-condition testing**, performed using controlled interval corruptions to evaluate numerical stability, warning visibility, and correction behavior.

**5. Duration sensitivity analysis**, performed to evaluate the effect of recording duration on frequency-domain interpretability and metric stability.

**6. Arrhythmia robustness and quality-control stress testing**, performed using selected arrhythmic recordings to evaluate numerical finiteness, warning behavior, and sensitivity to preprocessing under non-normal rhythm conditions.

A compact overview of validation phases, datasets, comparators, and primary evaluation objectives is provided in **Table 1**. Importantly, the MITDB analysis segments listed in Table 2 were treated exclusively as engineering robustness and quality-control stress-test material and were not incorporated into the primary HRV agreement analyses.

***Table 1.*** *Summary of validation phases, datasets, comparators, and primary evaluation objectives.*

| Validation Phase | Dataset / Data Source | Comparator / Reference | Primary Objective | Primary Endpoints |
|---|---|---|---|---|
| **Large-scale NeuroKit2 agreement validation** | **MIT-BIH Normal Sinus Rhythm Database (NSRDB)** and complementary PhysioNet RR datasets; clean normal-sinus- | NeuroKit2 (matched preprocessing and spectral settings) | Evaluate metric-level agreement under harmonized preprocessing and spectral conventions | Relative error, correlation, agreement, convention sensitivity |

| | rhythm recordings | | | |
|---|---|---|---|---|
| **Kubios benchmark subset comparison** | Kubios benchmark subset from quality-controlled recordings (50 exported → 44 reviewed files) | Kubios HRV software | Evaluate targeted benchmark agreement under matched settings | Relative error, correlation, median agreement, outlier sensitivity |
| **FFT and autoregressive (AR) spectral method evaluation** | Clean 10-min PhysioNet normal-sinus-rhythm recordings | HRV Studio Welch PSD reference; FFT and AR methods | Assess methodological consistency and support Welch PSD as the primary reporting method | LF/HF agreement, PSD consistency, AR-order sensitivity, finite-output rate |
| **Synthetic robustness and signal-condition testing** | Synthetic RR perturbations (missed beats, ectopy, jitter, dropouts, artifacts) | Clean same-recording baseline | Evaluate robustness under degraded signal conditions | Finite-output rate, warning visibility, preprocessing/correction behavior |
| **Duration sensitivity analysis** | PhysioNet normal-sinus-rhythm recordings truncated to 30 s, 60 s, 2 min, 3 min, 5 min, and 10 min | NeuroKit2 outputs at each matched recording duration | Evaluate duration dependence of cross-platform agreement and HRV metric stability | Relative error, duration stability, finite-output rate, warning behavior |
| **Arrhythmia robustness and quality-control stress testing** | **MIT-BIH Arrhythmia Database (MITDB)** selected recordings | No clinical comparator; robustness-focused evaluation | Evaluate numerical robustness and quality-control visibility under non-normal rhythms | Finite-output rate, warning behavior, preprocessing sensitivity, correction effects |

To improve reproducibility of the arrhythmia robustness experiments, the exact MITDB analysis cohort is summarized in Table 2. The robustness study comprised 12 predefined analysis segments derived from 8 unique MIT-BIH Arrhythmia Database recordings. These segments were intentionally selected to expose the preprocessing and quality-control pipeline to a diverse range of rhythm conditions, including predominantly normal rhythm, ventricular ectopy, supraventricular ectopy, mixed irregular rhythms, and noisy recordings. The objective was not to represent the overall distribution of arrhythmias within MITDB but rather to evaluate software robustness under heterogeneous rhythm conditions.

***Table 2.*** *MIT-BIH Arrhythmia Database analysis segments used for robustness and quality-control stress testing.*

| Segment ID | MITDB Record | Dominant rhythm / primary characteristic | Purpose in validation |
|---|---:|---|---|
| mitbih_101_0600_1200 | 101 | Mostly normal rhythm | Near-normal reference for QC behavior |
| mitbih_103_0000_0600 | 103 | Mostly normal rhythm | Near-normal reference for QC behavior |
| mitbih_103_0300_0900 | 103 | Mostly normal rhythm | Repeated normal segment to assess consistency |
| mitbih_105_0600_1200 | 105 | Noisy/problematic rhythm | Robustness to noisy recordings and QC warnings |
| mitbih_105_0900_1500 | 105 | Noisy/problematic rhythm | Robustness to noisy recordings and QC warnings |
| mitbih_106_0600_1200 | 106 | PVC/ectopic-heavy rhythm | Ventricular ectopy stress testing |
| mitbih_106_1200_1800 | 106 | PVC/ectopic-heavy rhythm | Ventricular ectopy stress testing |
| mitbih_200_0900_1500 | 200 | PVC/ectopic-heavy rhythm (including ventricular tachycardia episodes) | Severe ventricular arrhythmia robustness |
| mitbih_207_1200_1800 | 207 | Mixed irregular rhythm with ventricular and supraventricular ectopy | Mixed-arrhythmia QC stress testing |
| mitbih_208_0300_0900 | 208 | PVC/ectopic-heavy rhythm | High ectopic burden stress testing |
| mitbih_232_0900_1500 | 232 | Predominantly supraventricular ectopic rhythm | Supraventricular arrhythmia robustness |
| mitbih_232_1200_1800 | 232 | Predominantly supraventricular ectopic rhythm | Supraventricular arrhythmia robustness |

## Comparator Harmonization and Spectral Processing Conventions

To minimize implementation-dependent discrepancies during software comparison, comparator frameworks were harmonized using explicit preprocessing and spectral-processing conventions. Because frequency-domain HRV estimates are highly sensitive to preprocessing choices, interpolation, detrending, and spectral estimation parameters, the validation strategy sought to reduce avoidable disagreement arising from software configuration rather than physiological signal differences. Under the primary validation configuration, RR interval series were interpolated at **4 Hz**, Welch PSD estimation used a **120 s segment length**, **75% segment overlap**, and a **Hann window**, and equivalent preprocessing settings were adopted across comparator frameworks whenever practical. For the Kubios benchmark comparison, the detrending configuration was intentionally selected to match the processing settings of the exported Kubios reports rather than Kubios's default preprocessing workflow. Specifically, the exported benchmark reports were generated with detrending disabled ("no detrend"), and HRV Studio therefore used the same configuration to maximize methodological comparability and isolate differences attributable to software implementation rather than preprocessing choices. During subsequent validation review, sequence-level harmonization was identified as an additional requirement for cross-platform comparison. Although the same RR recordings were imported into both platforms, Kubios exported HRV metrics based on internally selected NN interval sequences (HRV.Data.RRs) rather than necessarily the complete imported interval vector. Therefore, final benchmark comparisons were repeated using the exact Kubios-selected NN interval sequences as input to HRV Studio, ensuring that both implementations operated on identical effective analysis segments before metric-level agreement was assessed.

For direct Kubios benchmarking, an explicit comparator-compatible spectral convention was used to reproduce the frequency-band definitions represented in the exported Kubios outputs. Under this configuration, $VLF$ was integrated over 0–0.04 Hz, $LF$ over 0.04–0.15 Hz, $HF$ over 0.15–0.40 Hz, and total power over 0–0.40 Hz. This comparator-specific convention was restricted to matched Kubios validation and was kept distinct from HRV Studio's standard physiological analysis mode, which uses mutually exclusive bands and excludes the DC component.

It should be noted, however, that Kubios Standard normally recommends the Smoothness Priors detrending method for routine short-term HRV analysis. Smoothness Priors effectively suppresses slow baseline drift before spectral estimation and generally provides more stable estimates of VLF and absolute spectral power. Although HRV Studio implements Smoothness Priors detrending based on the published Tarvainen formulation, exact implementation equivalence with Kubios could not be assumed because all internal processing details were not exposed. The no-detrend configuration used in the primary benchmark should therefore be interpreted as a comparator-matching strategy rather than the recommended preprocessing workflow for routine HRV analysis. Consequently, residual disagreement in the primary benchmark should be interpreted in the context of convention-sensitive spectral-processing differences rather than attributed solely to the underlying spectral estimation algorithms. Several implementation details were recognized as convention-sensitive and capable of materially affecting frequency-domain outputs, including handling of the direct-current (DC) component, detrending strategy, PSD integration behavior, and frequency-bin treatment near spectral-band boundaries. These factors predominantly influence $VLF$ and absolute spectral power, whereas normalized measures ($LF_{nu}$ and $HF_{nu}$) and the $LF/HF$ ratio are comparatively less sensitive to such preprocessing conventions. Accordingly, methodological differences arising from preprocessing and spectral conventions were interpreted separately from differences attributable to physiological signal content.

## 2.5 Statistical Analysis and Validation Endpoints

### Primary Endpoints

Validation performance was evaluated using complementary agreement, robustness, and quality-control endpoints selected to characterize both numerical behavior and interpretability of HRV outputs. For agreement validation, commonly used time-domain indices (RMSSD, SDNN, and pNN50) and nonlinear Poincaré descriptors (SD1 and SD2) were additionally evaluated alongside frequency-domain metrics to assess consistency across complementary HRV representations. The primary quantitative endpoint was **relative error**, computed to assess agreement between HRV Studio and comparator frameworks at the metric level:

$$\text{Relative error (\%)} = \left|\frac{HRV_{Studio} - Comparator}{Comparator}\right| \times 100 \tag{8}$$

where comparator values corresponded to matched-setting NeuroKit2, Kubios, or method-specific reference outputs depending on the validation phase. Correlation analysis was additionally performed to evaluate consistency of inter-recording metric structure between implementations. Because strong correlation alone does not imply numerical agreement, correlation metrics were interpreted alongside relative error distributions. Operational robustness was assessed using **finite-output rate**, defined as the proportion of analyses producing numerically finite outputs without computation failure. **Warning rate** was evaluated to characterize visibility of quality-control indicators under degraded or convention-sensitive conditions. Additional endpoints included **artifact-correction behavior**, assessed by comparing pre-correction and post-correction metric outputs; **PSD consistency diagnostics**, used to evaluate spectral plausibility and methodological stability; and **duration stability**, assessed by examining metric behavior across progressively truncated recording lengths.

#### Statistical Summaries

Because frequency-domain HRV comparisons can exhibit skewed error distributions and sensitivity to retained edge cases, robust summary statistics were prioritized. Median values were treated as the primary descriptive statistic for agreement analyses, particularly for the Kubios benchmark subset where a limited number of retained pathological or convention-sensitive recordings could disproportionately influence arithmetic means. Mean statistics were reported as secondary, outlier-sensitive summaries to provide additional descriptive context but were not used as primary headline evidence when distributions were strongly skewed. Metric-level interpretation emphasized agreement patterns rather than isolated summary values and considered both numerical agreement and methodological sensitivity. Comparisons were interpreted in a metric-specific manner, recognizing that some HRV measures are inherently more sensitive to preprocessing, detrending, interpolation, and PSD integration conventions than others.

### Interpretation Boundaries

Interpretation of validation results was intentionally restricted to claims supported by the completed evidence package. Frequency-domain metrics were interpreted in the context of their known sensitivity to preprocessing and spectral-analysis conventions. $VLF$ and absolute spectral power measures were evaluated particularly cautiously because of their sensitivity to detrending, DC handling, interpolation, PSD integration, recording duration, and frequency-bin treatment. Consequently, disagreement in these measures was considered potentially convention-sensitive rather than necessarily indicative of implementation error. The present validation focused on software behavior under explicitly harmonized preprocessing conditions.

Arrhythmic recordings were treated exclusively as engineering robustness and quality-control stress-test data and were not incorporated into the primary HRV agreement analyses.

# 3. Results

## 3.1 Large-Scale NeuroKit2 Agreement

Large-scale agreement analyses were performed using matched preprocessing and spectral-analysis settings to evaluate consistency between HRV Studio and NeuroKit2 across normal-sinus-rhythm recordings derived from PhysioNet resources. The five-minute analysis was treated as the primary short-term validation condition, consistent with established HRV assessment practice, while an extended ten-minute segment-linear analysis was performed as a secondary sensitivity assessment of recording-duration effects. The five-minute validation run included 15,179 recordings corresponding to 106,253 metric rows, whereas the ten-minute sensitivity analysis included 7,598 recordings and 53,186 metric rows. In the primary five-minute comparison, HRV Studio demonstrated strong agreement with NeuroKit2 for $LF$, $HF$, $LF/HF$, $LF_{nu}$, and $HF_{nu}$under matched preprocessing and the corrected standard frequency-band convention. Median relative errors were 1.35% for $LF$, 0.18% for $HF$, 13.57% for total power, 1.41% for $LF/HF$, 0.42% for $LF_{nu}$, 0.81% for $HF_{nu}$, and 37.79% for $VLF$. These findings indicate strong consistency for $LF$, $HF$, normalized, and ratio-based measures under harmonized settings, while confirming substantially greater convention sensitivity for $VLF$ and, to a lesser extent, total power.

The extended ten-minute sensitivity analysis demonstrated the same overall metric-dependent pattern, with generally lower disagreement for convention-sensitive spectral outputs. Median relative errors were 1.21% for $LF$, 0.11% for $HF$, 6.03% for total power, 1.25% for $LF/HF$, 0.37% for $LF_{nu}$, 0.72% for $HF_{nu}$, and 22.72% for $VLF$. Corresponding Pearson correlations were 0.9998 or higher for $LF$, $HF$, $LF/HF$, $LF_{nu}$, and $HF_{nu}$, 0.999 for total power, and 0.998 for VLF. Relative to the primary five-minute comparison, median $VLF$ error decreased from 37.79% to 22.72% and total-power error from 13.57% to 6.03%, whereas $LF$, $HF$, $LF/HF$, $LF_{nu}$, and $HF_{nu}$ remained comparatively stable across the two durations. The ten-minute analysis therefore serves as sensitivity evidence showing that increased recording duration can reduce disagreement for some convention-sensitive spectral outputs without eliminating their methodological sensitivity.

As a complementary analysis beyond the primary five-minute frequency-domain comparison, commonly used time-domain and nonlinear HRV indices were evaluated using the matched 10-minute segment-linear framework. As summarized in Table 3, HRV Studio demonstrated excellent agreement with NeuroKit2 for RMSSD, SDNN, and SD1, with median relative errors approaching zero and Pearson correlations of 1.000 under the evaluated implementation. pNN50 also demonstrated highly consistent agreement, with a median relative error of 0.135% and Pearson correlation approaching unity. SD2 exhibited slightly larger variability under the initial NeuroKit2 comparison framework (median relative error: 2.729%; Pearson r = 0.988), reflecting minor implementation-specific differences in Poincaré descriptor computation. SD2 also demonstrated near-identical agreement following harmonization of the calculation approach, with a median relative error of 0.053% and Pearson correlation of 0.99998. These results extend the validation beyond spectral measures and demonstrate consistent agreement across complementary time-domain and nonlinear representations of cardiac variability.

***Table 3.*** *Agreement of time-domain and nonlinear HRV indices between HRV Studio and NeuroKit2 under matched 10-minute segment-linear settings*

| Metric | Domain | Number of recordings | Median relative error (%) | Mean relative error (%) | Pearson correlation coefficient (r) |
|---|---|---|---|---|---|
| RMSSD | Time-domain | 7598 | 0 | 0 | 1 |
| SDNN | Time-domain | 7598 | 0 | 0 | 1 |
| pNN50 | Time-domain | 7364 | 0.135 | 0.137 | 1 |
| SD1 | Nonlinear | 7598 | 0 | ~0.000 | 1 |
| SD2 | Nonlinear | 7598 | 0.0525 | 0.1041 | 0.99998 |

***Note.*** *Agreement was evaluated using the same 10-minute segment-linear NeuroKit2 validation framework described in Section 3.1 (n = 7598 PhysioNet normal-sinus-rhythm recordings). Relative error was calculated as the absolute difference between HRV Studio and NeuroKit2 divided by the NeuroKit2 value. For pNN50, recordings with zero NeuroKit2 values were excluded from relative-error calculations. Small residual differences in pNN50 and SD2 reflect minor implementation-level calculation differences under the matched validation framework.*

Collectively, these findings support strong matched-setting agreement between HRV Studio and NeuroKit2 across multiple HRV domains, including time-domain, frequency-domain, and nonlinear indices. Frequency-domain metrics demonstrated particularly strong consistency under harmonized preprocessing and spectral conventions, whereas VLF and absolute-power estimates remained more sensitive to methodological choices involving detrending, spectral resolution, and low-frequency estimation.

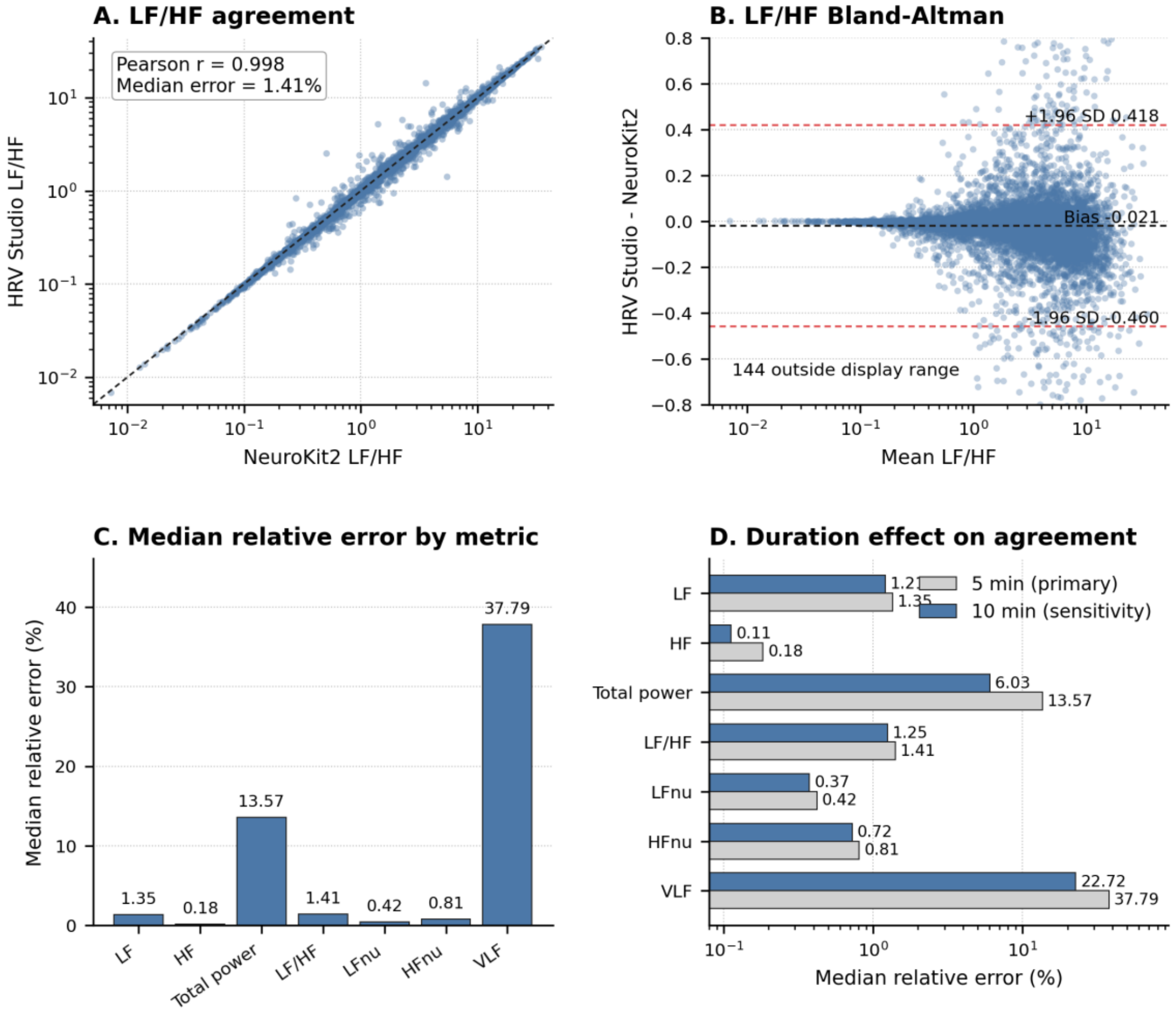

***Figure 3.*** *Agreement between HRV Studio and NeuroKit2 under the corrected standard frequency-band convention and matched preprocessing settings. (A) LF/HF agreement for the primary five-minute PhysioNet validation run. (B) Bland–Altman analysis for LF/HF in the primary five-minute validation. (C) Metric-wise median relative errors for the primary five-minute validation run. (D) Comparison of the primary five-minute results with the extended ten-minute sensitivity analysis, showing strong agreement for LF, HF, LF/HF, LFnu, and HFnu while VLF and total power remain more convention-sensitive.*

Inspection of early agreement discrepancies showed that residual differences between HRV Studio and NeuroKit2 were strongly influenced by preprocessing and PSD conventions rather than reflecting generalized disagreement in spectral structure. Earlier analyses performed under no-detrend conditions demonstrated substantial low-frequency sensitivity to direct-current (DC) handling and baseline treatment, particularly for $VLF$ and total power. Subsequent methodological review therefore motivated explicit separation of DC handling, frequency-band integration, and detrending conventions in the final implementation. Under the corrected standard HRV Studio convention, physiological frequency bands were treated as mutually exclusive and the DC component was excluded from standard band-power integration. The primary five-minute NeuroKit2 validation consequently demonstrated strong agreement for LF, $HF$, $LF/HF$, $LF_{nu}$, and $HF_{nu}$, while $VLF$ and total power retained greater convention sensitivity. The extended ten-minute sensitivity analysis reproduced the same overall metric-dependent pattern and showed reduced disagreement for several convention-sensitive spectral outputs, without changing the five-minute analysis as the primary short-term validation condition. These findings reinforce that cross-platform frequency-domain agreement depends not only on the spectral estimator itself, but also on explicit harmonization of detrending, band boundaries, DC treatment, interpolation, PSD integration, and recording duration.

## 3.2 Kubios Benchmark Subset

A targeted Kubios benchmark analysis was performed to assess agreement between HRV Studio and an established external HRV analysis platform under constrained matched-setting conditions. Because Kubios export workflows required manual report generation and subsequent parsing, the benchmark subset was intentionally treated as a focused comparison rather than a large-scale equivalence assessment. The benchmark workflow initially included 50 manually exported Kubios recordings, all of which were successfully discovered and parsed. Following matching to the validation framework, 48 recordings were retained for comparator analysis. Application of automated quality-control filtering reduced the dataset to 46 recordings, after which manual review was performed to identify retained pathological or technically unsuitable cases. The final manuscript-facing benchmark subset therefore consisted of 44 recordings (Figure 4A). Although the source benchmark recordings represented short-term HRV segments extending up to approximately 10 minutes, the effective interval sequences used by Kubios varied across recordings because the software internally selected NN interval samples for metric computation. The Kubios-selected sequences ranged from approximately 61 to 601 s (median $\approx$451 s). Sequence-level harmonization was therefore applied in the final benchmark by analyzing the exact Kubios-selected NN interval sequences in HRV Studio, ensuring that both platforms operated on equivalent effective input sequences. This staged filtering strategy was implemented to prioritize interpretable matched comparisons while preserving representative edge cases sufficient to evaluate robustness of agreement. Consequently, the final benchmark subset should be interpreted as a quality-controlled and manually reviewed comparison cohort rather than an exhaustive representation of all possible recording conditions.

### Metric-Level Agreement

Following sequence-level harmonization, frequency-domain agreement improved substantially compared with the initial non-harmonized benchmark (Figure 4B–D). For this direct external benchmark, HRV Studio used the Kubios-compatible frequency-band convention described in Section 2.4 so that the calculated $VLF$, $LF$, $HF$, and total-power outputs corresponded to the band definitions represented in the exported Kubios results. Using the exact Kubios-selected NN interval sequences as input to HRV Studio, median relative errors were 1.46% for $LF_{nu}$, 3.79% for $HF_{nu}$, 5.55% for $LF/HF$, 2.51% for total power, 3.49% for $LF$, 2.54% for $HF$, and 14.01% for $VLF$ (Figure 4B). Pearson correlations were correspondingly high across all evaluated frequency-domain metrics (r = 0.973–0.998), with $LF/HF$ demonstrating r = 0.982 and a median relative error of 5.55% (Figure 4C). Compared with the initial non-harmonized comparison, sequence matching reduced median relative error by 80.4% for $LF_{nu}$, 66.2% for $HF_{nu}$, 68.4% for $LF/HF$, 87.4% for total power, 82.9% for $LF$, 91.2% for $HF$, and 56.1% for $VLF$ (Figure 4D). These findings indicate that a substantial proportion of the apparent disagreement in the initial benchmark resulted from differences in the effective NN interval sequences analyzed by the two platforms rather than from differences in the underlying HRV metric implementations. After sequence harmonization, strong agreement was observed across most spectral metrics, although $VLF$ retained the largest residual relative error, consistent with the greater sensitivity of low-frequency spectral estimates to preprocessing and spectral-processing conventions. Additional agreement analyses were performed for commonly used time-domain and nonlinear HRV indices to extend the Kubios benchmark beyond frequency-domain measures. During validation review, it was identified that direct metric comparison could be influenced by differences in the effective NN interval sequences analyzed by each platform. Specifically, Kubios exported metrics based on its internally selected NN interval samples (HRV.Data.RRs), whereas initial HRV Studio calculations were performed on the complete imported interval sequence. To eliminate this segment-selection effect, the benchmark comparison was repeated using the exact Kubios-selected NN interval sequences as input for HRV Studio. Under these harmonized conditions, HRV Studio demonstrated near-identical agreement with Kubios for RMSSD, SDNN, pNN50, SD1, and SD2. Median relative errors were below 0.001% for RMSSD, SDNN, pNN50, and SD1, while SD2 showed a median relative error of 0.0705% with a Pearson correlation coefficient of 0.9999. These results indicate that the previously observed differences were attributable to sequence-selection mismatch rather than differences in metric implementation.

***Table 4.*** *Agreement of time-domain and nonlinear HRV indices between HRV Studio and Kubios after NN-sequence harmonization in the manually reviewed benchmark subset*

| **Metric** | **Domain** | **Number of recordings** | **Median relative error (%)** | **Mean relative error (%)** | **Pearson correlation coefficient (r)** |
|---|---|---|---|---|---|
| RMSSD | Time-domain | 44 | 0.00084 | 0.00567 | 1 |
| SDNN | Time-domain | 44 | 0.00004 | 0.00006 | 1 |
| pNN50 | Time-domain | 38 | 0.00034 | 0.00222 | 1 |
| SD1 | Nonlinear | 44 | 0.00091 | 0.00564 | 1 |
| SD2 | Nonlinear | 44 | 0.07049 | 0.3006 | 0.9999 |

***Note.*** *Agreement was evaluated using the final manually reviewed 44-recording Kubios benchmark subset described in Section 3.2. For time-domain and nonlinear validation, HRV Studio calculations were repeated using the exact NN interval sequences selected by Kubios (HRV.Data.RRs) to ensure identical input data between platforms. Relative error was calculated using Kubios values as reference. For pNN50, recordings with zero Kubios values were excluded from relative-error calculations because the denominator-based metric becomes undefined. The remaining small numerical differences reflect rounding and implementation-level calculation precision rather than differences in analyzed interval sequences.*

Taken together, the Kubios benchmark subset supports strong cross-platform consistency for time-domain and nonlinear indices after input-sequence harmonization. The results demonstrate that when identical NN interval sequences are analyzed, HRV Studio reproduces Kubios calculations for commonly used variability measures with negligible numerical deviation. In contrast, remaining differences observed in frequency-domain absolute-power measures should primarily be interpreted in the context of preprocessing conventions, detrending strategy, and spectral estimation methodology. Therefore, the Kubios benchmark provides complementary evidence supporting the reproducibility of HRV Studio across both conventional variability indices and convention-sensitive spectral measures under explicitly controlled analytical conditions.

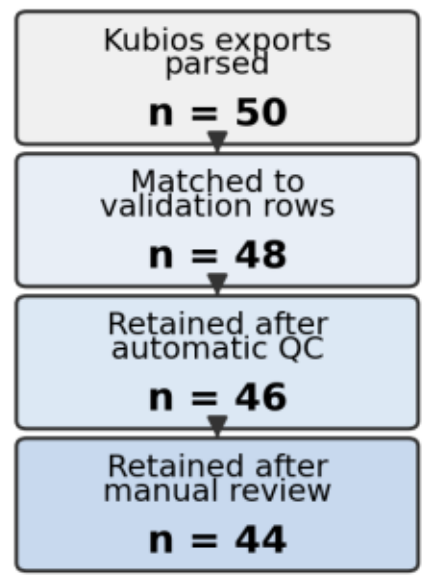

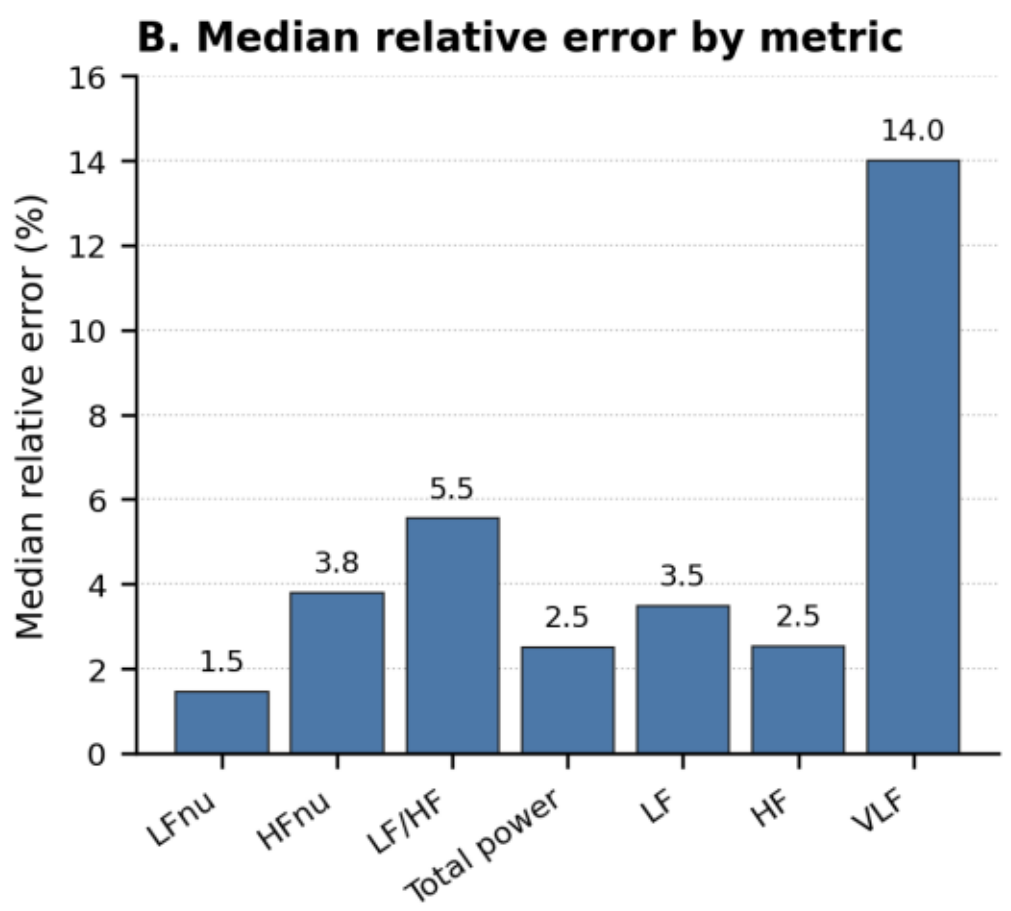

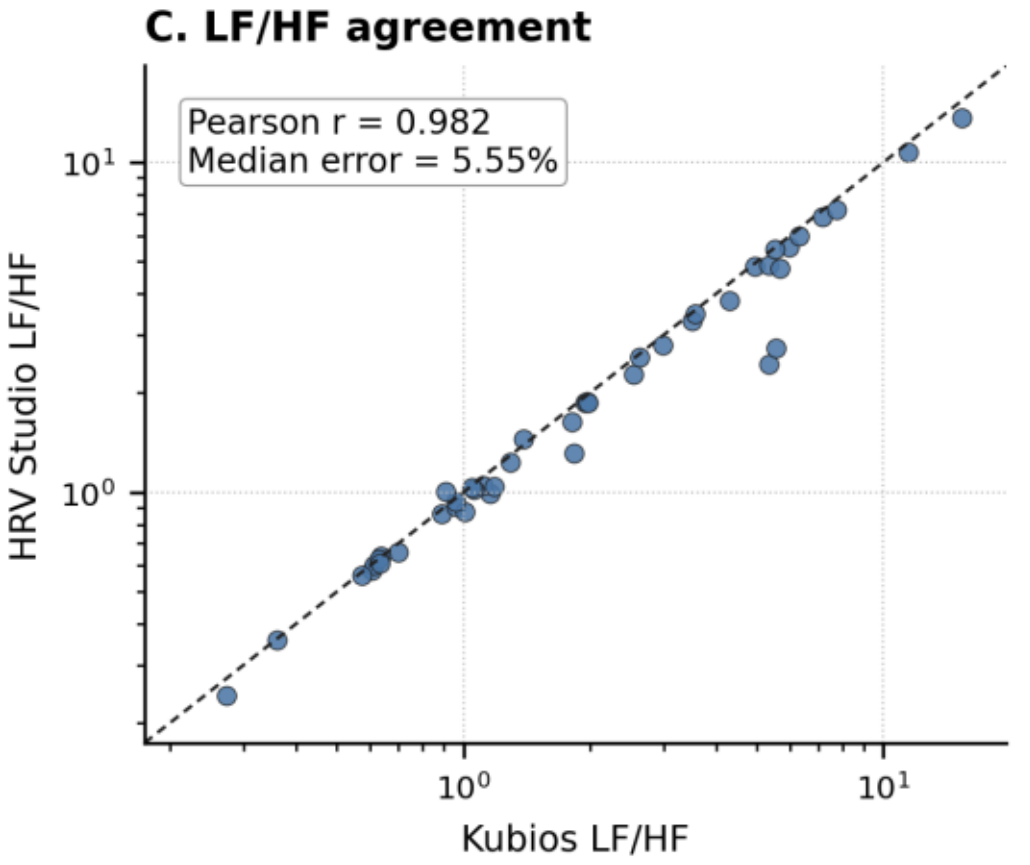

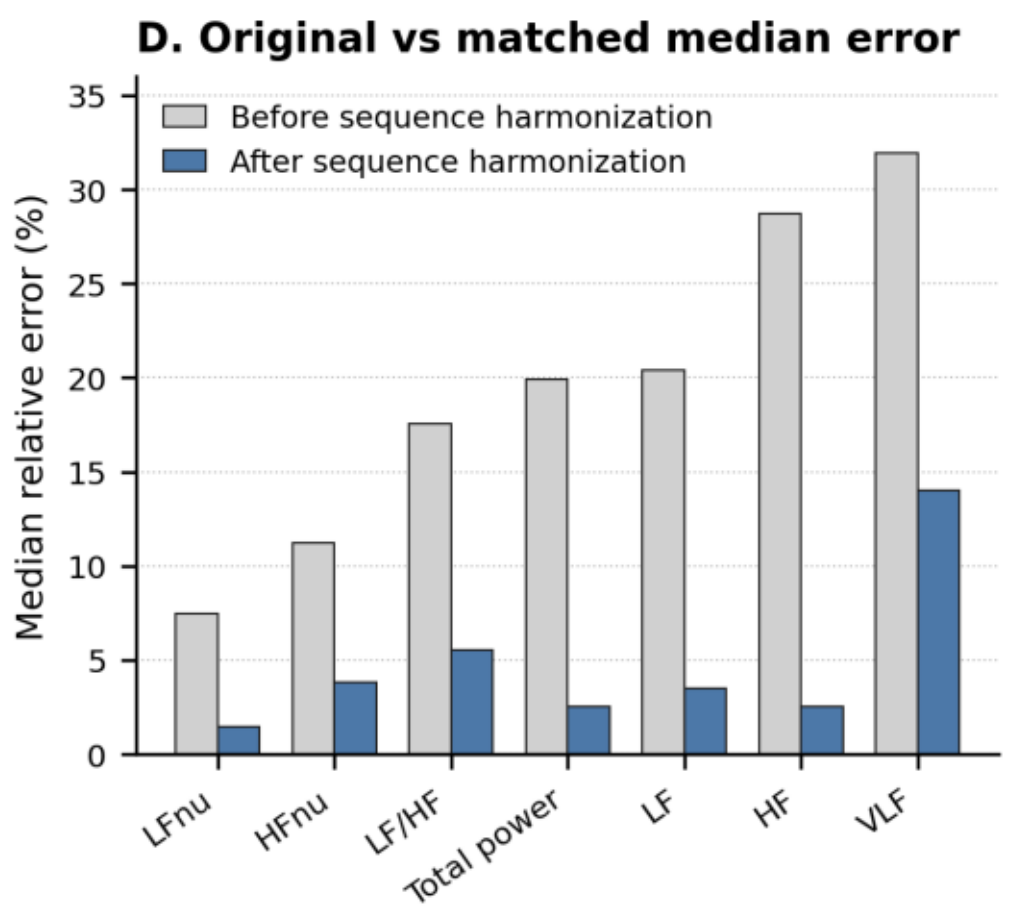

***Figure 4.*** *Kubios benchmark and effect of NN-sequence harmonization. (A) Filtering and quality-control workflow used to derive the final 44-recording benchmark subset from 50 parsed Kubios exports. (B) Median relative errors for frequency-domain metrics after sequence-level harmonization using the exact Kubios-selected NN interval sequences. (C) Agreement between HRV Studio and Kubios for LF/HF after sequence harmonization; the dashed line represents the line of identity. (D) Comparison of median relative errors before and after sequence harmonization, demonstrating the reduction in apparent cross-platform disagreement when equivalent effective NN interval sequences were analyzed. LF, low frequency; HF, high frequency; VLF, very low frequency; LF/HF, low-frequency to high-frequency power ratio; LFnu, low-frequency power in normalized units; HFnu, high-frequency power in normalized units.*

## 3.3 Smoothness Priors Sensitivity Analysis

To further investigate the influence of detrending on cross-platform agreement, a targeted sensitivity analysis was performed using a purposefully selected subset of ten recordings spanning representative clean/reference cases, previously identified VLF-sensitive recordings, manually reviewed outliers, review/control cases, and diagnostic control cases (Table 5) processed with nominal Smoothness Priors detrending ($\lambda = 500$) in both HRV Studio and Kubios. Unlike the primary Kubios benchmark presented above, which intentionally reproduced the no-detrend configuration recorded in the exported Kubios reports, this analysis evaluated agreement under matched Smoothness Priors preprocessing. In addition, all 10 recordings in this sensitivity analysis used the complete imported interval sequences in Kubios, with the Kubios-selected NN sequences matching the corresponding HRV Studio inputs; therefore, the comparison was not affected by effective sequence-selection differences. The purpose of this experiment was not to replace the primary benchmark, but rather to determine whether harmonizing detrending reduced residual differences between the two software packages.

***Table 5.*** *Recordings included in the Smoothness Priors sensitivity analysis.*

| Recording | Category | Selection rationale |
|---|---|---|
| CH001_nsr020_segment_000 | Clean/reference | Known clean high-agreement recording |
| CH002_nsr017_segment_023 | Clean/reference | Representative clean comparison recording |
| CH003_nsr030_segment_094 | Clean/reference | Representative clean comparison recording |
| CH004_nsr006_segment_000 | Clean/reference | Representative clean comparison recording |
| VLF002_nsr007_segment_062 | VLF-sensitive | Previously identified VLF-sensitive recording |
| VLF005_nsr008_segment_046 | VLF-sensitive | Previously identified VLF-sensitive recording |
| OUT005_nsr021_segment_071 | Outlier | Manually reviewed high-disagreement recording |
| OUT006_nsr015_segment_089 | Outlier | Manually reviewed high-disagreement recording |
| RC003_nsr002_segment_069 | Review/control | Previously manually reviewed control recording |
| OUT001_nsr009_segment_012 | Diagnostic control | Recording used in the original DC/baseline handling investigation |

***Note.*** *The subset was selected to span representative agreement conditions observed during the primary Kubios benchmark, including clean/reference recordings, VLF-sensitive cases, manually reviewed outliers, review/control cases, and diagnostic control recordings. The purpose of the subset was to evaluate the influence of matched Smoothness Priors preprocessing across a range of representative signal characteristics rather than to provide a statistically representative sample of the complete benchmark dataset.*

Inspection of the Kubios version used in this study revealed that Welch spectral estimation was not available as a selectable analysis method. Instead, Kubios exported separate FFT- and autoregressive (AR)-based frequency-domain results. Consequently, estimator-aware comparisons were performed between Kubios FFT and HRV Studio FFT, and between Kubios AR and HRV Studio AR. As in the primary Kubios benchmark, the Kubios-compatible frequency-band convention was retained to ensure that corresponding spectral outputs represented equivalent exported band definitions across the two platforms. Both software packages were configured using nominally matched preprocessing parameters, including Smoothness Priors detrending (λ = 500), 4 Hz interpolation, and identical frequency-band definitions. Kubios FFT exports additionally reported a 120-s analysis window with 75% overlap, whereas HRV Studio FFT used whole-signal spectral estimation. Moreover, the Kubios FFT window function was not exposed in either the graphical interface or exported reports, preventing exact implementation matching.

The resulting agreement differed substantially between spectral estimation methods (Table 6). AR-based frequency-domain metrics demonstrated consistently close agreement across all evaluated measures, with median relative errors of 1.40% (VLF), 1.39% (LF), 3.18% (HF), 1.74% (total power), 4.38% (LF/HF), 1.16% ($LF_{nu}$), and 2.14% ($HF_{nu}$). Correlation coefficients were correspondingly high, with Pearson's r ranging from 0.989 to 0.999 across all AR metrics. In contrast, FFT-based agreement was more variable. Median relative errors ranged from 4.10% for LFnu to 18.41% for LF, with VLF also showing substantial disagreement at 17.44%; Pearson correlations remained high for most metrics but were notably lower for VLF (r = 0.586). These findings indicate that, even under nominally matched

Smoothness Priors preprocessing, estimator-specific implementation differences contribute more substantially to cross-platform variability than detrending alone. Specifically, agreement depended strongly on the spectral estimation method employed. The consistently close AR agreement suggests that independently implemented autoregressive spectral estimation can produce highly comparable HRV metrics, whereas the larger FFT discrepancies likely reflect remaining implementation differences, including spectral windowing and FFT processing details that could not be fully harmonized because the corresponding Kubios implementation parameters were not publicly exposed. Accordingly, this experiment is best interpreted as a targeted sensitivity analysis that complements, rather than replaces, the primary Kubios benchmark.

***Table 6.*** *Agreement between HRV Studio and Kubios under matched Smoothness Priors preprocessing (λ = 500) for the 10-recording sensitivity subset.*

| **Metric** | **FFT Median Relative Error (%)** | **FFT Pearson *r*** | **FFT Spearman *ρ*** | **AR Median Relative Error (%)** | **AR Pearson *r*** | **AR Spearman *ρ*** |
|---|---|---|---|---|---|---|
| VLF | 17.44 | 0.586 | 0.624 | 1.4 | 0.997 | 1 |
| LF | 18.41 | 0.906 | 0.842 | 1.39 | 0.999 | 1 |
| HF | 6.62 | 0.988 | 0.964 | 3.18 | 0.989 | 0.964 |
| Total Power | 15.78 | 0.921 | 0.927 | 1.74 | 0.996 | 0.988 |
| LF/HF | 14.83 | 0.989 | 0.976 | 4.38 | 0.996 | 0.988 |
| $LF_{nu}$ | 4.1 | 0.989 | 0.988 | 1.16 | 0.989 | 0.988 |
| $HF_{nu}$ | 11.48 | 0.989 | 0.964 | 2.14 | 0.989 | 0.964 |

***Note.*** *Median relative errors and correlation coefficients were calculated from matched HRV Studio–Kubios comparisons using nominal Smoothness Priors detrending (λ = 500). FFT comparisons were performed between Kubios FFT and HRV Studio FFT, whereas AR comparisons were performed between Kubios AR and HRV Studio AR. Kubios Welch estimates were not available in the software version used for this study. Although nominal preprocessing parameters were matched, exact implementation equivalence could not be established because the Kubios FFT window function was not exposed in the software interface or exported reports.*

## 3.4 Frequency-Domain Method Checks (FFT/AR)

### Welch versus NeuroKit2

Frequency-domain method checks were performed to evaluate methodological consistency across spectral estimation approaches and to determine the most defensible primary reporting method within the HRV Studio framework. Consistent with the broader validation design, Welch PSD estimation was treated as the primary spectral method, while FFT and autoregressive (AR) approaches were evaluated as secondary analyses. Under matched preprocessing and spectral settings, HRV Studio Welch PSD showed strong agreement with NeuroKit2 Welch estimates, supporting its use as the principal frequency-domain reporting approach. In particular, Welch-derived estimates demonstrated strong correspondence across clean ten-minute PhysioNet recordings (r = 0.987), while the aggregate median relative error remained influenced by convention-sensitive frequency-domain measures (13.34%; Figure 5A). This pattern indicates strong preservation of inter-recording spectral structure despite metric-dependent differences in absolute agreement. Because Welch PSD exhibited stable numerical behavior together with strong matched-setting agreement, it was selected as the primary interpretation framework for subsequent validation analyses. Accordingly, FFT and AR outputs were interpreted primarily as methodological checks rather than interchangeable replacements for Welch-derived estimates.

### FFT and AR Comparisons

FFT and autoregressive spectral analyses were evaluated to assess methodological robustness and identify convention-sensitive behavior that could influence HRV interpretation. Both methods produced numerically finite outputs across all evaluated recordings (100/100 files), thus showing operational stability within the implemented analysis framework. Despite numerical finiteness, FFT-based estimates displayed substantial sensitivity to preprocessing conventions, particularly detrending and direct-current (DC) handling. PSD-area consistency diagnostics revealed pronounced mismatch between integrated FFT spectral area and signal variance under the evaluated no-detrend convention, with a median PSD-to-variance ratio of 96.757. Correspondingly, FFT instability flags were observed in 100/100 recordings, reflecting sensitivity to baseline treatment and low-frequency spectral inflation rather than computational instability (Figure 5B). Autoregressive spectral estimation indicated comparatively stronger PSD consistency, with a median PSD-area ratio of 1.000, but retained sensitivity to model specification. Although instability flags were uncommon (3/100 recordings), 25/100 recordings exhibited variability greater than 50% across tested AR model orders, indicating nontrivial order dependence in spectral estimates (Figure 5C). These findings support a method-dependent interpretation of frequency-domain outputs. FFT estimates remained particularly sensitive to detrending and DC conventions, whereas AR estimates showed stable but order-dependent behavior. Consequently, Welch PSD was retained as the primary reporting method in the present validation framework because it demonstrated the strongest combination of comparator agreement, methodological stability, and interpretability under matched settings. Within HRV Studio, an AR model order of 16 is used as the default configuration for short-term HRV analysis under the standard 4 Hz interpolation framework and served as the reference setting throughout the present study.

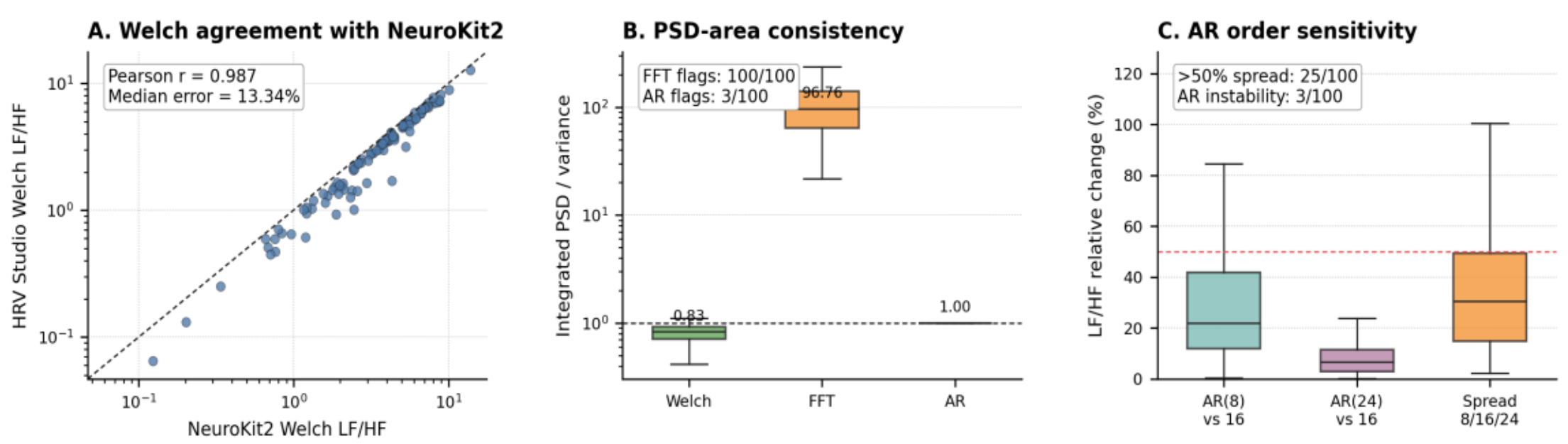

***Figure 5.*** *Comparison of spectral estimation methods in HRV Studio under the corrected standard frequency-band convention. (A) Agreement between HRV Studio Welch PSD and NeuroKit2 under matched preprocessing and spectral settings. (B) PSD-area consistency across Welch, FFT, and autoregressive (AR) estimation, illustrating pronounced FFT sensitivity under the evaluated no-detrend baseline. (C) AR model-order sensitivity, with 25/100 recordings showing >50% spread across tested orders and 3/100 triggering AR instability flags. These findings support Welch PSD as the primary reporting method within the present validation framework.*

## 3.5 Synthetic Robustness and QC Behavior

Synthetic robustness analyses were performed using controlled RR interval perturbations designed to emulate common signal degradations, including missed beats, ectopic short–long patterns, Gaussian timing jitter, short signal dropouts, and isolated extreme artifacts (Figure 6A–B). These analyses were intended to evaluate numerical stability, quality-control visibility, and sensitivity to preprocessing under degraded signal conditions. Across the larger robustness evaluation, HRV Studio maintained finite numerical outputs in all tested conditions (60/60 cases) both before and after preprocessing-based correction, demonstrating stable computational behavior despite substantial perturbation burden (Figure 6C). Quality-control warnings were also consistently surfaced, with diagnostic or warning indicators present in 60/60 corrected cases, supporting transparent identification of degraded or potentially unreliable recordings. Artifact correction materially affected both short-term variability and frequency-domain measures, including RMSSD, $LF$, $HF$, $LF/HF$, and total power, demonstrating the sensitivity of HRV estimates to interval corruption and preprocessing choices (Figure 6D). In contrast, metrics such as SDNN demonstrated comparatively greater stability under correction. These findings support the operational robustness of HRV Studio under controlled perturbation scenarios while emphasizing the importance of quality-control interpretation. Specifically, finite output was not assumed to imply physiological validity, and warning behavior was interpreted as a screening aid rather than confirmation of signal reliability.

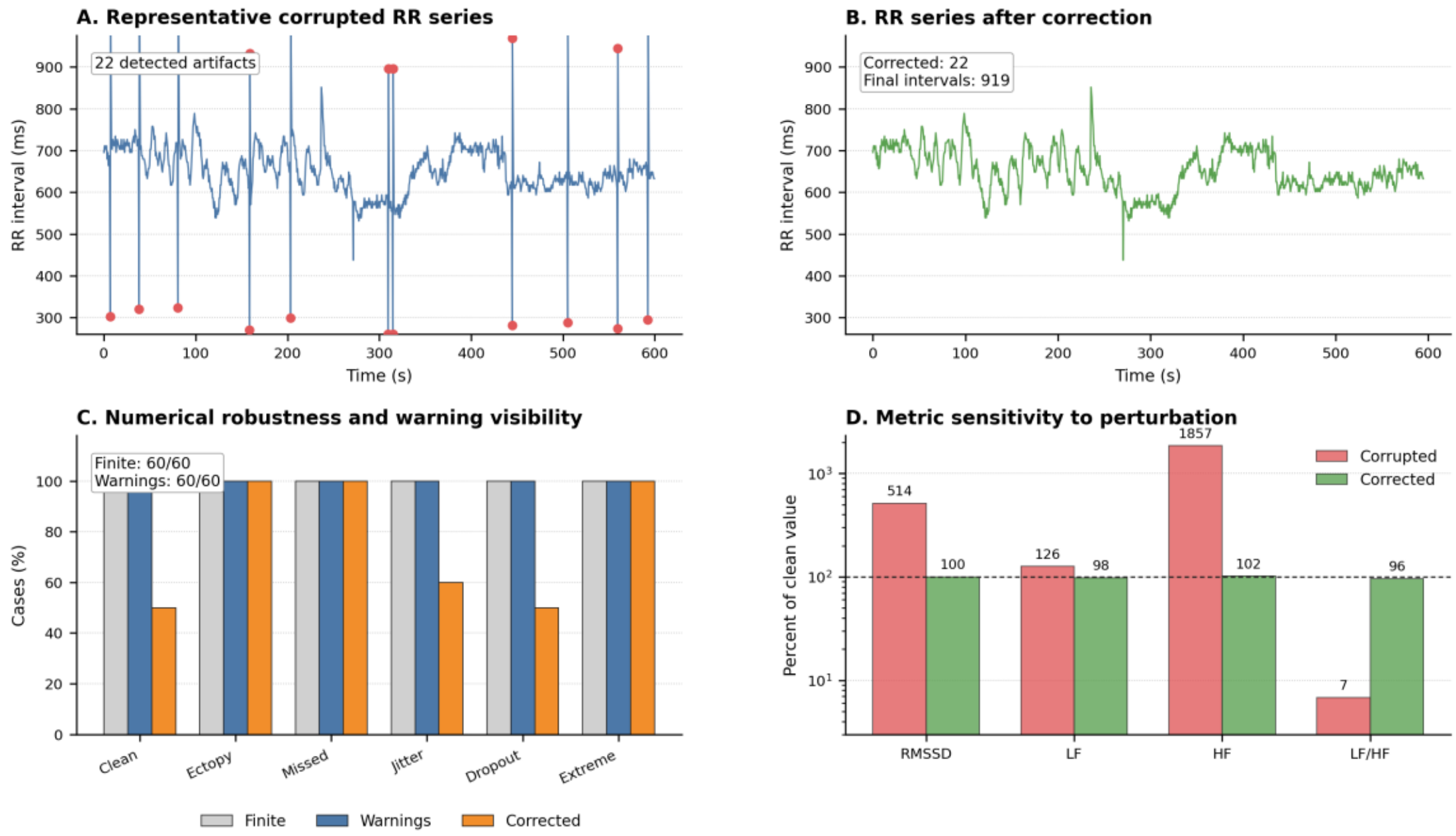

*Figure 6. Robustness of HRV Studio under controlled signal perturbations using the corrected standard analysis convention. (A) Representative corrupted RR interval series generated through controlled perturbation. (B) The same signal following preprocessing and correction. (C) Numerical robustness and quality-control warning visibility across perturbation scenarios; finite outputs and warnings were present in all 60 evaluated cases. (D) Representative effects of corruption and correction on RMSSD and selected frequency-domain metrics. The results demonstrate stable computation and transparent quality-control behavior under degraded signal conditions while emphasizing that finite output does not necessarily imply physiological validity.*

## 3.6 Duration Sensitivity

Duration sensitivity was evaluated by truncating the same clean RR recordings to 30 s, 60 s, 2 min, 3 min, 5 min, and 10 min and comparing HRV Studio outputs with NeuroKit2 outputs computed from the same duration-matched truncated RR segment. Unless otherwise specified, relative error at each duration was calculated as the absolute difference between the HRV Studio and NeuroKit2 values divided by the absolute NeuroKit2 value at that same duration. It should be noted that the five-minute duration results reported here originate from the dedicated duration-sensitivity experiment, in which shorter segments were truncated from the same clean recordings used for the extended ten-minute NeuroKit2 sensitivity analysis described in Section 3.1, whereas the five-minute values reported in Section 3.1 correspond to the independent large-scale five-minute NeuroKit2 agreement validation cohort. Accordingly, the two five-minute estimates reflect distinct experimental configurations and are not directly comparable.

Although HRV Studio maintained finite numerical outputs at all evaluated durations, frequency-domain agreement remained strongly dependent on both recording length and metric type (Figure 7A–C). $LF/HF$ demonstrated its greatest disagreement at 30 s (median relative error, 60.29%), followed by 27.66% at 60 s and 28.61% at 2 min. Agreement improved at longer durations, with median relative errors of 19.20% at 3 min, 11.98% at 5 min, and 12.57% at 10 min. $VLF$ exhibited substantially greater and non-monotonic sensitivity to duration. A reliable $VLF$ comparison was not available at 30 s; median relative error was 100% at 60 s, 41.86% at 2 min, 32.25% at 3 min, 33.71% at 5 min, and 38.75% at 10 min (Figure 7B). Thus, increasing recording duration substantially improved the stability of several conventional spectral measures relative to ultra-short recordings, but did not produce monotonic convergence for $VLF$. These findings support cautious interpretation of frequency-domain outputs below five minutes and identify five minutes as the practical minimum short-term duration within the present framework. $VLF$, however, remained convention-sensitive even at five- and ten-minute durations and should therefore be interpreted cautiously regardless of nominal short-term recording length.

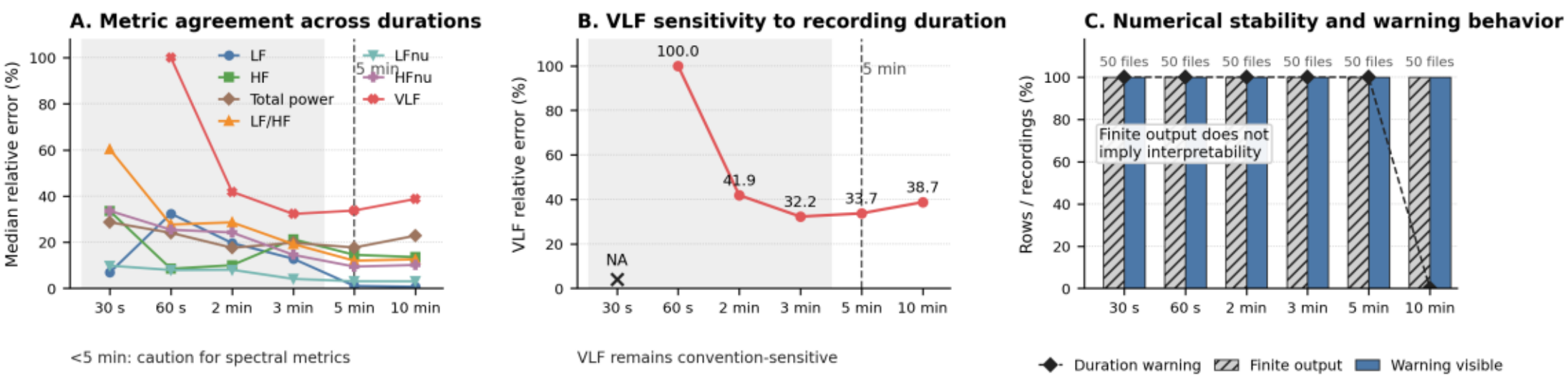

*Figure 7. Recording-duration sensitivity and warning behavior. (A) Median relative error between HRV Studio and NeuroKit2 for frequency-domain metrics, with both platforms evaluated on the same duration-matched truncated RR segments at 30 s, 60 s, 2 min, 3 min, 5 min, and 10 min. (B) VLF-specific median relative error relative to matched-duration NeuroKit2 outputs; the 30-s VLF value is unavailable because both platforms returned zero VLF power, resulting in a zero reference denominator. (C) Finite-output and warning rates across durations, distinguishing numerical computability from interpretability warnings.*

## 3.7 MIT-BIH Arrhythmia Robustness/QC Stress Testing

The predefined MITDB analysis cohort summarized in Table 2, comprising 12 analysis segments derived from 8 unique MIT-BIH recordings, was used to evaluate the numerical robustness and quality-control behavior of HRV Studio under non-normal rhythm conditions (Figure 8A–C). These analyses were designed as engineering stress tests of the preprocessing and quality-control pipeline rather than as agreement or clinical validation studies. Across all 12 predefined analysis segments, HRV Studio maintained finite outputs before and after preprocessing-based correction (12/12), demonstrating stable computational performance despite substantial rhythm irregularity (Figure 8D). Quality-control visibility was also consistent, with warnings present in 12/12 corrected recordings, supporting transparent identification of irregular or potentially unreliable conditions. Preprocessing and correction effects were most pronounced in recordings with substantial ectopy or rhythm irregularity, affecting both conventional variability measures such as RMSSD and frequency-domain outputs, whereas near-normal rhythm segments showed comparatively smaller changes. Collectively, these findings demonstrate that HRV Studio maintains stable numerical behavior while providing consistent quality-control visibility across heterogeneous arrhythmia conditions. As an engineering stress test, this evaluation demonstrates the robustness of the preprocessing and quality-control pipeline under challenging rhythm conditions that commonly disrupt automated HRV workflows.

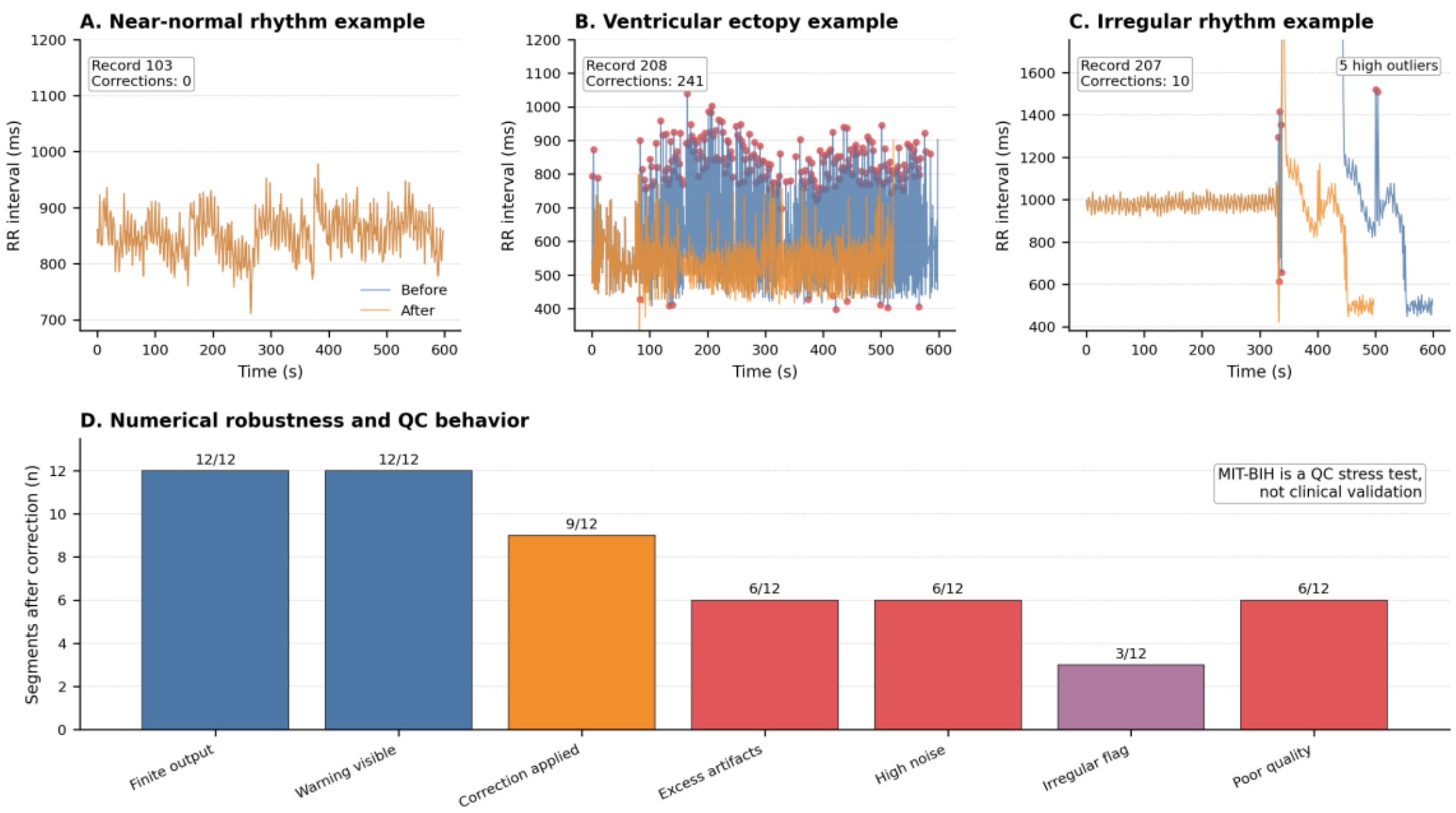

***Figure 8.*** *Arrhythmia robustness and quality-control behavior under non-normal rhythm conditions using the corrected standard analysis convention. (A) Representative near-normal rhythm segment. (B) Ventricular ectopy–dominant rhythm example illustrating extensive preprocessing under irregular beat structure. (C) Highly irregular rhythm example used as a quality-control stress-test condition. (D) Summary of numerical robustness and quality-control behavior across the 12 selected MIT-BIH segments; finite outputs and visible warnings were present in all cases, while correction and quality-control flags varied by rhythm condition. Results are interpreted strictly as engineering robustness and quality-control stress testing rather than clinical validation.*

# 4. Discussion

The present validation framework provides evidence that HRV Studio can support transparent and quality-control-aware HRV research workflows under explicitly defined preprocessing and spectral conventions. Across large-scale PhysioNet-derived normal-sinus-rhythm recordings, the primary five-minute comparison demonstrated strong agreement with NeuroKit2 under matched settings for $LF$, $HF$, $LF/HF$, $LF_{nu}$, and $HF_{nu}$, whereas $VLF$ and total power remained more sensitive to frequency-band and spectral-processing conventions. Agreement improved when preprocessing and spectral conventions were harmonized, highlighting the importance of methodological comparability in HRV software validation. NeuroKit2 was selected as the primary large-scale comparator because it provides an open, reproducible, and inspectable implementation capable of large-scale automated validation under harmonized settings. In contrast, Kubios benchmarking was necessarily constrained by manual export, report generation, and review workflows, limiting the scale of practical comparison. Consequently, NeuroKit2 served as the primary agreement framework, whereas Kubios provided complementary external benchmarking under carefully controlled conditions. The extended ten-minute NeuroKit2 analysis was used as a secondary sensitivity assessment and reproduced the same overall metric-dependent pattern, with lower disagreement for several convention-sensitive spectral outputs. This secondary result does not alter the five-minute analysis as the primary short-term validation condition. The separate duration-sensitivity experiment likewise showed substantially greater instability in ultra-short recordings and persistent, non-monotonic $VLF$ sensitivity across short-term durations. Taken together, these findings suggest that, when preprocessing and PSD assumptions are aligned, HRV Studio can provide reproducible and interpretable HRV outputs suitable for research-oriented applications.

Validation performance varied across HRV metrics and experimental conditions, but the sequence-harmonized Kubios benchmark substantially changed the pattern observed in the initial comparison. After identical effective NN interval sequences were analyzed by both platforms, strong agreement was observed not only for $LF/HF$, $LF_{nu}$, and $HF_{nu}$, but also for $LF$, $HF$, and total power, with median relative errors below 6% for these measures. $VLF$ retained the largest residual relative error (14.01%), indicating comparatively greater sensitivity to low-frequency preprocessing and spectral-processing conventions. Together with the large-scale NeuroKit2 results, these findings indicate that apparent metric-specific disagreement can arise from both analytical conventions and differences in the effective interval sequences submitted to computation. Consequently, cross-platform HRV comparisons should control the analyzed NN sequence as well as detrending, PSD estimation, frequency-band definitions, and recording duration before residual differences are attributed to metric implementation.

Beyond frequency-domain measures, the additional validation of time-domain and nonlinear indices demonstrated consistent agreement across complementary representations of HRV variability. In the large-scale NeuroKit2 comparison, RMSSD, SDNN, and SD1 exhibited near-identical agreement under matched preprocessing conditions, with median relative errors approaching zero and correlation coefficients of 1.000. pNN50 also demonstrated highly consistent agreement, although its relative-error interpretation was affected by recordings with very small or zero reference values. Following harmonization of the SD2 calculation approach, SD2 likewise demonstrated near-identical agreement, further supporting consistency of the nonlinear metric implementation. The targeted Kubios benchmark further demonstrated near-identical cross-platform consistency for these metrics after NN-sequence harmonization. Importantly, the sequence-level audit showed that the apparent discrepancies in the initial Kubios comparison were attributable primarily to differences in the effective NN interval sequences analyzed rather than to disagreement in the deterministic metric definitions. Once identical NN interval sequences were analyzed, HRV Studio reproduced Kubios time-domain and nonlinear calculations with negligible numerical differences, confirming consistency of the underlying metric implementations.

A central methodological finding of the present validation study was therefore that meaningful cross-platform HRV comparison depends on harmonization at two complementary levels: the effective NN interval sequence submitted to analysis and the subsequent preprocessing and spectral-analysis conventions. In the primary Kubios benchmark, both platforms received the same source RR recordings, but Kubios calculated its exported metrics from internally selected NN interval samples that were shorter than the complete imported sequences in 31 of the 44 recordings. Repeating the benchmark using these exact Kubios-selected sequences substantially reduced frequency-domain disagreement, including reductions in median relative error from 17.56% to 5.55% for $LF/HF$ and from 31.92% to 14.01% for $VLF$. This demonstrates that providing identical source recordings alone does not necessarily guarantee equivalent analytical input across software platforms and that effective sequence selection should be explicitly controlled during cross-platform validation. The final implementation further distinguishes between two spectral-band conventions according to analytical purpose. Standard HRV Studio analysis uses mutually exclusive physiological bands with the direct-current component excluded, whereas direct Kubios benchmarking uses an explicit comparator-compatible convention reproducing the frequency-band definitions represented in the exported Kubios outputs. This separation prevents external-software compatibility requirements from being conflated with the default physiological reporting convention and makes the selected spectral definition an explicit component of the reproducible analysis workflow.

After removal of this sequence-selection confounder, the remaining frequency-domain differences further emphasized the role of explicit methodological conventions. The primary Kubios benchmark intentionally reproduced the no-detrend configuration recorded in the exported Kubios reports to maximize methodological comparability under those explicit settings. To further evaluate the influence of detrending, an additional Smoothness Priors sensitivity analysis was performed using matched preprocessing in both HRV Studio and Kubios. All 10 recordings included in this sensitivity analysis were confirmed to use identical effective interval sequences in both platforms, excluding NN-sequence selection as an explanation for the observed differences. Under matched Smoothness Priors preprocessing, AR-derived metrics exhibited substantially closer agreement than FFT-derived metrics, with the remaining differences occurring primarily for VLF and absolute spectral powers. Because sequence alignment and detrending were both controlled in this experiment, the persistence of larger FFT discrepancies supports the contribution of estimator-specific spectral-processing differences. In particular, the HRV Studio FFT implementation used whole-signal spectral estimation, whereas the Kubios exports reported 120-s FFT windows with 75% overlap; the exact Kubios FFT window function was not exposed in the available exports. These differences therefore represent plausible contributors to the residual FFT disagreement, although their individual effects were not isolated experimentally. Differences in detrending strategy, direct-current (DC) handling, and PSD integration behavior can also materially influence frequency-domain outputs, especially $VLF$ and total power. Earlier no-detrend analyses under the legacy and comparator-oriented conventions revealed inflation of low-frequency spectral content attributable to DC-related baseline effects; this observation motivated both the segment-linear matched-setting validation framework and the explicit exclusion of the DC component from HRV Studio's final standard physiological band convention. Accordingly, transparent reporting of interpolation, detrending, windowing, segment behavior, effective NN-sequence selection, and PSD integration conventions remains essential for reproducible HRV analysis and meaningful cross-platform comparison.

From an engineering perspective, the no-detrend configuration should primarily be regarded as a comparator-matching option rather than the preferred preprocessing strategy for routine HRV analysis. HRV Studio provides multiple detrending methods, including None (no detrending), Constant detrending (mean/DC removal), Linear detrending (global linear baseline removal), and Smoothness Priors detrending based on the published Tarvainen formulation (default $\lambda = 500$). These options support different analytical objectives, with Smoothness Priors providing the closest methodological counterpart to Kubios's default preprocessing for routine short-term HRV analysis, consistent with the AR-versus-FFT sensitivity pattern noted above.

$VLF$ and absolute spectral power should be interpreted within the context of both detrending strategy and spectral estimation methodology. However, the sequence-harmonized Kubios results indicate that absolute LF, HF, and total power can also achieve strong cross-platform agreement when the effective NN sequence and relevant processing settings are appropriately controlled. VLF remained comparatively more sensitive than the other evaluated spectral measures, while normalized measures ($LF_{nu}$, $HF_{nu}$) and $LF/HF$ continued to demonstrate robust agreement across the validation framework.

Recording-duration analyses provided an additional caution regarding short-term spectral interpretation. Ultra-short recordings showed substantially greater disagreement across several frequency-domain measures, supporting the continued use of approximately five minutes as a practical minimum for conventional short-term spectral analysis. However, $VLF$ did not demonstrate monotonic improvement with increasing duration: median relative error remained 33.71% at five minutes and 38.75% at ten minutes in the duration-sensitivity experiment. Thus, extending a short-term recording alone does not resolve $VLF$ instability, and low-frequency interpretation remains dependent on spectral resolution, preprocessing, and band-integration conventions.

The observed AR model-order sensitivity also has practical implications for software configuration. HRV Studio uses a default AR order of 16 for short-term HRV analysis under the standard 4 Hz interpolation framework. This choice is consistent with previous methodological work recommending an order of at least 16 for spectral analysis of short tachogram segments [35] and with the default AR configuration used in Kubios. Although another investigation reported statistically similar normalized spectral indices across AR orders from 9 to 25 [36], the present results showed that LF/HF could remain strongly order-dependent in a subset of recordings. Accordingly, order 16 should be regarded as a defensible starting configuration rather than a universally optimal value. When AR-derived VLF or LF/HF results are central to interpretation, users should assess robustness across a limited set of nearby orders, such as 16–20, and avoid interpreting conclusions that change materially with model order [37]. Welch PSD remains the preferred primary reporting method within HRV Studio because it does not require AR-order selection and demonstrated stronger overall methodological stability in the present validation.

HRV Studio was designed to integrate quality-control visibility into the HRV analysis workflow rather than treat preprocessing as an opaque step. Across synthetic perturbation scenarios and arrhythmia-focused stress testing, warning systems consistently surfaced conditions associated with degraded interpretability, including excessive correction burden, rhythm irregularity, short recording duration, and unstable low-frequency estimation. Importantly, finite numerical output was not interpreted as evidence of physiological validity. Stable computation under degraded signal conditions indicates operational robustness but does not guarantee that resulting HRV metrics are physiologically meaningful or suitable for interpretation. Accordingly, warning behavior is intended to function as a screening and interpretability aid rather than a substitute for investigator review.

Beyond automated quality-control diagnostics, HRV Studio also supports investigator-supervised review and correction of interval series following automated waveform processing. Although waveform-derived interval extraction was not treated as a primary validation endpoint in the present study, automated R-peak detection remains a recognized source of downstream HRV error, particularly in recordings containing motion artifact, baseline wander, ectopic beats, or other signal abnormalities. HRV Studio therefore complements automated detection with interactive beat-editing tools that allow users to visually inspect interval series, insert, delete, move, or interpolate individual beats, perform immediate reanalysis, reverse modifications through undo functionality, and maintain an auditable editing history. These capabilities do not replace automated preprocessing or constitute validation of waveform-derived interval extraction. Rather, they provide a transparent human-in-the-loop workflow that helps bridge the practical gap between raw physiological waveforms and high-quality interval series before downstream HRV interpretation.

The sequence-harmonized Kubios comparison and the additional Smoothness Priors sensitivity analysis provide complementary external benchmarking under two distinct methodological conditions. The primary benchmark demonstrated strong cross-platform agreement across time-domain, most frequency-domain, and nonlinear metrics once equivalent effective NN interval sequences were analyzed. The Smoothness Priors analysis further showed that, even when both sequence selection and detrending were matched, spectral-estimator-specific differences could persist, particularly for FFT-derived low-frequency and absolute-power measures. Taken together, these results support strong reproducibility for RMSSD, SDNN, pNN50, SD1, SD2, LF, HF, total power, $LF_{nu}$, $HF_{nu}$, and $LF/HF$ under appropriately harmonized conditions, while identifying $VLF$ as the spectral measure requiring the greatest methodological caution. The findings also demonstrate that cross-platform disagreement should not be attributed to software implementation before effective sequence selection and analytical conventions have been systematically controlled.

Similarly, analyses using recordings from the MIT-BIH Arrhythmia Database (MITDB) were intentionally restricted to robustness and quality-control stress testing. Numerical stability and warning visibility under non-normal rhythm conditions demonstrate operational robustness but do not constitute clinical validation of HRV analysis under arrhythmic conditions or evidence that HRV metrics derived from such recordings are physiologically reliable without appropriate clinical interpretation.

## Limitations

Several limitations should be considered when interpreting the present findings. First, the Kubios benchmark subset was relatively small, manually exported, and manually curated, limiting generalizability and increasing sensitivity to retained edge cases. This limitation is particularly relevant to relative-error statistics for metrics such as pNN50 when reference values are very small or zero. The Kubios comparison depended on manually exported reports and parsed outputs, which limited access to certain internal processing details, particularly for frequency-domain analysis. For the final benchmark, effective NN interval sequences were explicitly harmonized across both conventional HRV and frequency-domain comparisons using the Kubios-selected interval samples, thereby removing sequence-selection mismatch as a source of residual cross-platform disagreement. Nevertheless, some internal spectral-processing details could not be fully reconstructed from the exported Kubios outputs. In particular, although the exported settings documented FFT segmentation parameters, certain implementation details such as the exact FFT window function were not exposed. Residual frequency-domain differences may therefore reflect preprocessing, spectral integration, segmentation/windowing, and other estimator-specific implementation choices that could not be completely harmonized. In addition, direct Kubios frequency-domain comparisons used a comparator-specific band convention matching the exported Kubios definitions, whereas HRV Studio's standard analysis mode uses mutually exclusive physiological bands with the DC component excluded. Accordingly, Kubios benchmark values for $VLF$ and total power should be interpreted as comparator-matched outputs and should not be assumed to be numerically interchangeable with standard-mode HRV Studio $VLF$ and total-power estimates. Consequently, the Kubios results are best interpreted as targeted external benchmarking under sequence-harmonized conditions rather than as evidence of complete software equivalence across all HRV measures. The additional Smoothness Priors sensitivity analysis complemented this benchmark by evaluating matched detrending conditions but was intentionally performed on a targeted subset of recordings and should therefore be interpreted as a methodological sensitivity analysis rather than a large-scale validation. Importantly, all 10 recordings in this sensitivity analysis were confirmed to use identical effective interval sequences in both platforms, excluding sequence-selection mismatch as an explanation for its residual spectral differences. Also, the present results do not establish performance across broader populations, wearable-device artifacts, ambulatory monitoring environments, or clinically heterogeneous cohorts and may not fully represent all real-world acquisition conditions. Second, $VLF$ and, more generally, absolute spectral power measures remain sensitive to preprocessing conventions, detrending behavior, spectral estimation methodology, and recording duration, limiting direct comparability across implementations. Although sequence harmonization produced strong Kubios agreement for $LF$, $HF$, and total power in the present benchmark, these results should not be generalized to differently configured analysis pipelines without equivalent methodological harmonization. Third, short-duration recordings demonstrated substantially greater instability, particularly for low-frequency spectral metrics, supporting cautious interpretation below five minutes. Fourth, results from the MIT-BIH Arrhythmia Database (MITDB) should not be interpreted as clinical validation, as arrhythmia recordings were used solely for robustness and quality-control stress testing. Finally, FFT- and AR-based spectral estimates demonstrated method dependence, and residual variability attributable to PSD conventions and implementation-specific spectral processing remained despite harmonized settings. Consequently, frequency-domain agreement should be interpreted in the context of explicit preprocessing and methodological assumptions.

# 5. Conclusion

This study presented the validation of HRV Studio, a Python/PyQt6-based platform for heart rate variability analysis designed to emphasize transparency, reproducibility, and quality-control visibility. Validation was performed through a staged framework including large-scale comparison with NeuroKit2, targeted benchmarking against Kubios, spectral-method evaluation, synthetic robustness testing, duration sensitivity analysis, and arrhythmia-focused quality-control stress testing. Across these validation phases, HRV Studio demonstrated consistent agreement across complementary HRV domains, including time-domain, frequency-domain, and nonlinear indices, when preprocessing and analytical conventions were explicitly defined and reported. Collectively, these results demonstrate that HRV Studio can support quality-control-aware HRV research workflows under transparent and reproducible analysis conditions.

Agreement was strongly dependent on methodological harmonization, particularly for frequency-domain analyses. Frequency-domain metrics demonstrated strong cross-platform reproducibility under harmonized conditions, including $LF$, $HF$, total power, normalized spectral measures, and $LF/HF$, while commonly used time-domain indices such as RMSSD, SDNN, and pNN50, together with nonlinear Poincaré descriptors, further supported the consistency of HRV Studio across complementary representations of cardiac variability. In contrast, $VLF$ and total/absolute spectral power measures remained comparatively more sensitive to frequency-band definitions, DC handling, detrending behavior, PSD integration choices, recording duration, and other methodological factors, supporting their more cautious interpretation across differently configured analysis pipelines. These findings reinforce the importance of transparent preprocessing and explicit reporting of spectral-analysis assumptions in HRV software validation. They further demonstrate the importance of distinguishing standard physiological analysis conventions from comparator-specific settings used for external software benchmarking. Future work will focus on larger-scale external benchmarking, expanded Kubios comparisons, evaluation using prospective device-derived datasets, and assessment within broader clinical cohorts. Such studies will further extend the validation framework presented here while preserving the emphasis on transparent, quality-control-aware HRV analysis.